\documentclass[aps,prb,twocolumn,superscriptaddress,showpacs,longbibliography]{revtex4-1}
\usepackage[colorlinks=true,citecolor=blue,linkcolor=blue,breaklinks=true]{hyperref}
\usepackage{amssymb}
\usepackage{graphicx}
\usepackage{amsmath}
\usepackage{comment}
\usepackage{mathtools}
\usepackage[export]{adjustbox}
\usepackage{epsfig}
\usepackage{times}
\usepackage{ulem}
\usepackage{xcolor}
\usepackage{subfigure}
\usepackage{setspace}
\usepackage{bm}% bold math
\usepackage{calc}
\usepackage{natbib}
\begin{document}

\newcommand{\fm}[1]{\textcolor{black}{#1}}
\newcommand {\ba} {\ensuremath{b^\dagger}}
\newcommand {\Ma} {\ensuremath{M^\dagger}}
\newcommand {\psia} {\ensuremath{\psi^\dagger}}
\newcommand {\psita} {\ensuremath{\tilde{\psi}^\dagger}}
\newcommand{\lp} {\ensuremath{{\lambda '}}}
\newcommand{\A} {\ensuremath{{\bf A}}}
\newcommand{\Q} {\ensuremath{{\bf Q}}}
\newcommand{\kk} {\ensuremath{{\bf k}}}
\newcommand{\qq} {\ensuremath{{\bf q}}}
\newcommand{\kp} {\ensuremath{{\bf k'}}}
\newcommand{\rr} {\ensuremath{{\bf r}}}
\newcommand{\rp} {\ensuremath{{\bf r'}}}
\newcommand {\ep} {\ensuremath{\epsilon}}
\newcommand{\nbr} {\ensuremath{\langle ij \rangle}}
\newcommand {\no} {\nonumber}
\newcommand{\up} {\ensuremath{\uparrow}}
\newcommand{\dn} {\ensuremath{\downarrow}}
\newcommand{\rcol} {\textcolor{red}}
\newcommand{\jp}[1]{\textcolor{blue}{#1}}

\begin{abstract}
The effects of different forms of weak measurements on the nature of the measurement induced phase transition are theoretically studied in hybrid random quantum circuits of qubits. We use a combination of entanglement measures, ancilla purification dynamics, and a transfer matrix approach to compute the critical exponents, the effective central charge, and the multifractal spectrum of the measurement induced transitions. We compare weak measurements with an infinite number of discrete outcomes to a protocol with only a pair of outcomes and find that to within our numerical accuracy the universal critical properties are unaffected by the weak measurement protocols and are consistent with the universality class found for strong projective measurements.
\end{abstract}

%\title{Weak Measurements in Quantum Circuits}% Force line breaks with \\
%\title{Critical properties of weak measurement induced entanglement transitions in Quantum Circuits}
\title{Critical Properties of Weak Measurement Induced Phase Transitions in Random Quantum Circuits}

\author{Kemal Aziz}
 \affiliation{Department of Physics and Astronomy, Center for Materials Theory, Rutgers University, Piscataway, NJ 08854, USA}
 \author{Ahana Chakraborty}
 \affiliation{Department of Physics and Astronomy, Center for Materials Theory, Rutgers University, Piscataway, NJ 08854, USA}
 \author{J. H. Pixley}
 \affiliation{Department of Physics and Astronomy, Center for Materials Theory, Rutgers University, Piscataway, NJ 08854, USA}
 \affiliation{Center for Computational Quantum Physics, Flatiron Institute, 162 5th Avenue, New York, NY 10010}

%\collaboration{CLEO Collaboration}%\noaffiliation

\date{\today}% It is always \today, today,
             %  but any date may be explicitly specified

%\keywords{Suggested keywords}%Use showkeys class option if keyword
                              %display desired
\maketitle

%\tableofcontents

\section {Introduction}

Measures of quantum entanglement, such as Reyni entanglement entropies \cite{RenyiReviewPlenio,EntanglementKitaev,EntanglementReviewHorodecki,MBLSerbyn,Calabrese_2005}, provide critical insights into a variety of equilibrium and non-equilibrium properties in quantum many-body systems. 
In hybrid quantum circuits \cite{FisherReview,Potter2022}, where generic unitary dynamics competes with disentangling random local measurements, 
a measurement induced phase transition (MIPT) takes place in the structure of the entanglement \cite{NahumPRX,FisherMIPTPRB,FisherMIPTPRB2,AltmanPRB,AltmanPRLMIPT,SmithPRBMIPT,MIPT_Ising}. The  critical point  has been shown to be Lorentz invariant\cite{FisherCFT_2021, AltmanPRB,NahumPRX, PhysRevB.100.134203} and can be  described by a logarithmic conformal field theory (log-CFT)\cite{JianMIPT, Aidan_PRL} 
in one-dimensional chains  (though several perturbations, such as long-range gates~\cite{NormanPRL} or static measurement profiles that are disordered~\cite{staticdisorder} or quasiperiodic~\cite{Shkolnik-Gazit-2023}, can dramatically modify this critical behavior to no longer be Lorentz invariant). The properties of the log-CFT depends crucially on the quantum nature of the problem and can be numerically investigated through a bulk or boundary transfer matrix approach~\cite{Aidan_PRL,kumar2023boundary}. In both stabilizer circuits and qudits of local Hilbert space size $q\rightarrow \infty$, the MIPT does not have multifractal correlations~\cite{Aidan_PRL,li2021statistical} that reflect the discrete nature of Clifford gates and the classical nature of the percolation problem, respectively. In Haar random gates on qubit chains on the other hand, the multi-fractal correlations are strongly pronounced providing a qualitative distinction between these problems\cite{LUDWIG1987687,Aidan_PRL}.  

A monitored circuit with weak, as opposed to strongly projective, measurements that extract only partial information from the system provide a much more versatile implementation of an open quantum system. Microscopically, a generic example of a means to implement a weak measurement is via an ancilla degree of freedom (e.g. a qubit) coupled to the system~\cite{nielsen2010quantum}. 
Upon measurement of the ancilla, the state of the system is updated based on the outcome of the ancilla measurement. Such measurements are generically given by a measurement strength $J$ which can be used to tune between no observation ($J=0$) and projective measurements ($J\rightarrow \infty$) of some system observable. At intermediate strengths, only partial information of the systems observable that is being weakly measured can be obtained. Unlike 
projective measurements, which are described by projection-valued measures, weak measurements are described by positive operator valued measures (POVMs)\cite{nielsen2010quantum}. 
The size of POVMs 
may be larger than the dimension of the Hilbert space of the qubit being measured  (because its elements are not necessarily orthogonal), which leads to the possibility of making simultaneous measurements of non-commuting observables~\cite{Ochoa_2018,PhysRevLett.60.1351}. Putting such weak measurements into a hybrid random quantum circuit, as depicted in Fig.~\ref{fig:schematicmodel}, thus represents an interesting class of models that have more entanglement than its strongly projective counterpart. 
Previous work has shown that the MIPT can remain in the presence of weak measurements provided the strength of the measurement is not too weak\cite{AltmanPRB,WeakSchomerus,doggen2023ancilla,PhysRevLett.125.210602}. 

For the  MIPT in the $q\rightarrow \infty$ limit 
it is 
%commonly believed 
expected 
%found
that the nature of the universality class is independent of the type of  random local measurements being strongly projective or weak~\cite{Barratt-2022}. However,  detailed studies of the role of different types of local measurements and different types of gates, e.g. Haar random~\cite{Barratt-2022,doggen2023ancilla} versus free fermions~\cite{KellsRomito-2023,nehraDganit-2024} evolution, remains a central problem.
An aspect of this question is directly investigated in this work through a detailed numerical study of the effects of weak measurements on the MIPT in Haar random hybrid quantum circuits. 
We compare and contrast two distinct models for a weak measurement; one representing an infinite number of measurement outcomes, with one that has a binary outcome.
We focus on dual Haar random gates as this reduces the error in the analysis of the transfer matrix as shown in Ref.\onlinecite{Aidan_PRL}. The universality class of the MIPT in Haar and dual Haar models are expected to be the same ~\cite{Aidan_PRL,Zabalo-2020} and we further verify this by computing the remaining (straightforwardly accessible) unknown exponents with strong projective measurements in Appendix \ref{sec:HDU}.
Using purification based probes on an auxiliary ancilla qubit we provide unbiased estimates of the location of the MIPTs as a function of the measurement strength. 
Across the MIPT we use finite size scaling of the entanglement entropy of an ancilla qubit, and the mutual information between a pair of ancilla qubits to extract several critical exponents and show
that the transition remains Lorentz invariant. This allows us to apply the bulk transfer matrix construction of Ref.~\cite{Aidan_PRL} to study the universal  properties of the log-CFT.

To formulate the free energy of the log-CFT~\cite{Aidan_PRL} based on the entropy of the measurement record and further the Lyapunov spectra of the transfer matrix, we find that it is essential to have discrete, not continuous, measurement outcomes.
To make this construction explicit we
review 
a model for  weak measurements  with continuous outcomes to then show how we  can  effectively ``bin'' the measurement outcomes to a width $\epsilon$ allowing us to construct  a  model with a discrete but infinite number of measurement outcomes. 
Using this approach we compute the leading Lyapunov spectrum of the transfer matrix  to extract the effective central charge of the log-CFT, the typical scaling dimension of the order parameter, and the multifractal spectrum of the order parameter. To summarize, all of the typical critical exponents agree well with the strong projective case while the leading multifractal exponent that we have computed does have a slightly larger deviation. Taken together, our results strongly suggest that the universality class of the MIPT is unaffected by going from strong to weak measurements and they are ultimately described by the same log-CFT.

The remainder of this paper is organized as follows: In Sec.~\ref{sec:model}, we introduce the models of the monitored circuits including the Haar dual unitary entangling gates and three different protocols for weak measurements. In Sec.~\ref{sec:phasediagram}, we compute the phase diagram of the entanglement transition as a function of the measurement strength $J$ and measurement rate. We provide numerical evidence of Lorentz invariance at the transition and compute several critical exponents.
In Sec.~\ref{sec:TM}, we study the log-CFT governing the transition through a transfer-matrix based approach and calculate the effective central charge, the typical scaling
dimension of the order parameter, and the multifractal spectrum. We conclude and provide outlook for future works in Sec.~\ref{conclusion}. In Appendix~\ref{sec:HDU} we compute critical exponents of the strongly projective dual unitary Haar model, while in Appendix~\ref{sec:AppendixB} and ~\ref{sec:AppendixC} we provide additional details on the models with an infinite number of measurement outcomes. Lastly in Appendix~\ref{app:ceff} we show that introducing a discrete measurement outcome does not affect our estimate of the critical properties.

\section {Models}\label{sec:model}
In the following section, we will describe the models we use for the entangling unitary gates and the various forms of weak measurements.
We consider a class of hybrid random quantum circuits, depicted in Fig.~\ref{fig:schematicmodel}(a), consisting of a chain of qubits where the unitary dynamics is generated by the  entangling gates 
(blue squares) that we take to be either randomly drawn from the Haar random or dual Haar random distribution (defined below). Random local measurements are applied to each site with probability $p$ and the red circles in Fig.~\ref{fig:schematicmodel}(a) denote where a measurement has taken place. We define one timestep as one layer of gates, followed by one layer of measurements. 
With increasing $p$, for both strong projective measurements~\cite{NahumPRX,Zabalo-2020} and (certain strengths of) weak measurements~\cite{WeakSchomerus}, this hybrid circuit exhibits a MIPT in its entanglement structure.
In the following, we explore the effects of different forms of weak measurements on the nature of the universality class of the MIPT. 
In particular, we consider three models that differ in the nature (or protocol) of the weak measurement we apply. The first two models involve weak measurements with an infinite number of outcomes that can be continuous or discrete. The third model involves only two outcomes but a ``softened'' projection operator.
\subsection{Entangling Unitary gates} \label{model:gate}
We aim to study the most generic quantum many body circuit of qubits. At the level of the unitary gates this is achieved by sampling each gate randomly from the Haar distribution of random U$(4)$ matrices. 
Importantly, previous work~\cite{Aidan_PRL,kumar2023boundary} has shown that we can restrict this generic gate set to a smaller subset of ``dual-unitary" Haar (HDU) gates~\cite{ProsenDU,LamacraftDU,LamacraftDU2},
which are unitary along the space and time direction, and still probe the same transition while obtaining more accurate numerical results for the free energy of the log-CFT (explained in more detail in Sec.~\ref{sec:TM} and see Refs.~\cite{Aidan_PRL,kumar2023boundary}).  
In Appendix~\ref{sec:HDU}, we explore this universality class further to provide additional evidence beyond Ref.~\onlinecite{Aidan_PRL}  that the strong projective MIPT in random HDU and random Haar circuits are within the same universality class.  
As a result,
for the majority of the paper,
we use two site HDU qubit gates between neighboring sites (unless otherwise specified) that are given by, 
\begin{equation}
U = e^{i \phi}(U_{+} \otimes U_{-})\cdot V[\theta] \cdot (V_{-} \otimes V_{+})
\end{equation}
where $\phi, \theta \in \mathbb{R}$ are chosen randomly from $[0,\pi)$ and $U_{\pm}, V_{\pm} \in SU(2)$, are randomly chosen from the Haar measure, and
\begin{equation}
V[\theta] = \exp\left[-i\left(\frac{{\pi}}{4}\sigma_{x} \otimes \sigma_{x} + \frac{{\pi}}{4}\sigma_{y} \otimes \sigma_{y} + \frac{{\pi}}{4}\theta\sigma_{z} \otimes \sigma_{z}\right)\right].
\end{equation}
Here, $\sigma^x, \sigma^y,\sigma^z$ are the Pauli spin-1/2 matrices.

\subsection{Models for Weak Measurement} \label{model:measurement}
\begin{figure}
\includegraphics[scale=0.32]{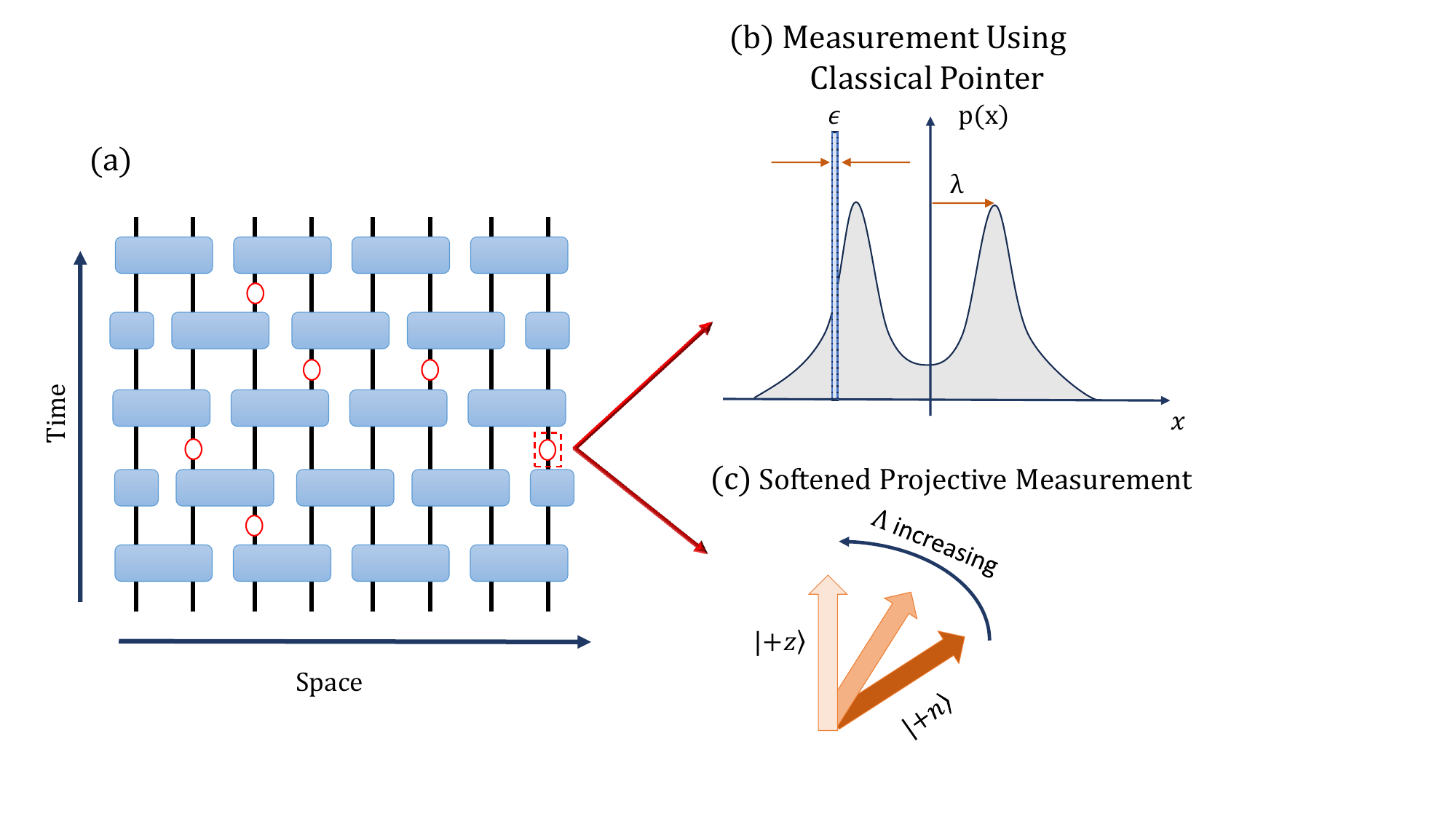}
\caption{{\bf Circuit model and different measurement protocols}:
(a) We consider models of a hybrid quantum circuit with a bricklayer geometry acting on a chain of length $L$ of spin-1/2 qubits with periodic boundary conditions. The two-site unitary gates (blue squares) are interspersed with weak measurements at every site with probability $p$ (red circles denote the location that a measurement has taken place).  (b) In both the models coupled to the classical pointer (see Secs.~\ref{measureCont} and~\ref{measureDiscrete}), $p(x)$ denotes the probability distribution of the location of a classical Gaussian pointer coupled to a qubit. 
The outcome of a readout operation on the pointer in the DGPM is shown by the blue region of width $\epsilon$. The outcome of the readout operation on the pointer in the CGPM is denoted by a continuous position x. The strength of the measurement depends on the ratio of the separation between the two Gaussian peaks ($\lambda$) and the width of the Gaussian distribution $(\Delta)$. (c) The measurement set-up for the Softened Projective Measurement Model (SPMM). For a single qubit oriented along the $|+n\rangle$ direction in the Bloch sphere, intermediate values of $\Lambda$ interpolate between $|+z\rangle$ and $|+n\rangle$. 
}
\label{fig:schematicmodel}
\end{figure}
To implement a weak measurement locally in the circuit, we use the generic description of the von Neumann model \cite{von2018mathematical}, where the system is first entangled with an ancilla locally, and then a projective measurement is performed on the ancilla. We consider three weak measurement models. In the first two models, we consider a measuring device with a ``classical pointer''~\cite{WeakSchomerus} having canonically conjugate position $\hat{x}$ and momentum $\hat{q}$ operators, satisfying $[\hat{x},\hat{q}] = i$ (in units with $\hbar=1$). 
The classical pointer interacts with a qubit of the system for a time $\delta t$ with a coupling strength $\lambda$, which  is then followed by a readout of the pointer position. The readout location $x_{o}$ is a continuous variable with an infinite number of outcomes.
This readout operation on the pointer updates the state of the qubit in the system either partially or fully, being controlled by $\lambda$. This implements a weak measurement on the qubit with varying strength of measurement. We first provide a review  of the continuous outcome model in Sec.~\ref{measureCont}. However, we are unable (at present) to form a transfer matrix description of this MIPT  due to the infinitesimal Born probabilities associated with it. To overcome this limitation, we  bin the measurement outcomes $x_o$ to a small window $\epsilon$ around that point to form a discrete weak measurement model, that is described in detail in Sec.~\ref{measureDiscrete}. 

We also find it interesting to contrast and compare these models with a weak measurement model with only a pair of outcomes. 
In subsection \ref{measureSoft}, we consider a weak measurement model obtained from softening the projective measurement of the $z-$component of the qubit, which implements a weak measurement with two outcomes.

\subsubsection{Continuous Gaussian pointer}\label{measureCont}
In this section, we 
review a well known model for 
a weak measurement model where the system is coupled to a classical pointer that can be measured continuously at position $x$\cite{WeakSchomerus}. We dub this the continuous Gaussian pointer model (CGPM). This will set the stage in the following section to define a similar model with discrete measurement outcomes. 

The wavefunction of the pointer is initialized in a Gaussian state $|\phi(x_{c} = 0) \rangle$ of width $\Delta$ centered at $x_{c} = 0$, which can be expanded in terms of the position basis states $|x\rangle$ as: 
\begin{equation}
|\phi(x_{c} = 0) \rangle = \frac{1}{\sqrt{\Delta}}\int \limits_{-\infty}^{\infty} a(x) |x\rangle dx. 
\end{equation}
The corresponding probability amplitudes are chosen from a squared Gaussian distribution, namely $|a(x)|^{2}=[G_{\Delta}(x)]^{2}$,
and
\begin{equation}
G_{\Delta}(x) = \frac{e^{\frac{-x^{2}}{2 \Delta^{2}}}}{\pi^{1/4} \Delta^{1/4}} 
\end{equation}
is a Gaussian distribution of width $\Delta$ centered at $x=0$. 
On the other hand, the system is initialized in a state $\vert \psi \rangle$ which can be written in terms of the $2^{L}$ product states $|e_{i}\rangle$ spanning the Hilbert space of $L$ qubits with corresponding expansion coefficients $c_i$s as,
\begin{equation}
\vert \psi \rangle = \sum_{i=1}^{2^{L}}c_{i}|e_{i}\rangle.
\end{equation}
Hence the system-pointer combined initial state at $t=0$ is given by, 
\begin{equation}
|\Psi(t=0)\rangle = |\psi \rangle \otimes |\phi(x_{c} = 0 ) \rangle.
\end{equation}
At the start of each measurement operation on a system qubit ($t=0$), we couple the system with the measuring device with a tunable coupling strength $\lambda$ via the Hamiltonian $\hat{H}_{\mathrm{int}}$, 
\begin{equation}
    \hat{H}_{\mathrm{int}}=\lambda \Theta(t)\Theta(\delta t-t) \sigma_{z}^{(j)} \otimes \hat{q}.
    \label{eq:Hint}
\end{equation}
Here the pointer interacts with the $j^{th}$ site of the system where z-component of the spin is to be measured and $\Theta(x)$ denotes the Heaviside step function. The system-pointer interaction is turned on for an interval of time $\delta t$ imposed by the theta function. 
During this time interval, the system and pointer jointly evolve under the unitary operator,
\begin{equation}
\hat{U}_{\mathrm{int}}(\delta t) = e^{-i\hat H_{\mathrm{int}}\delta t} = \Pi_{+}^{(j)}\otimes e^{-i\lambda {\delta t} \hat{q}} + \Pi_{-}^{(j)}\otimes e^{i\lambda {\delta t} \hat{q}},
\end{equation}
where 

\begin{equation}
    \Pi_{\pm}^{(j)} = [(\mathbb{I} \pm \sigma_{z}^{(j)})/2 ]\otimes_{i \not= j} \mathbb{I}^{(i)}
    \label{eqn:spinproject}
\end{equation}
projects the $z-$component of the $j^{th}$ spin onto the spin-up, or the spin-down state. We set $\delta t =1$ without any loss of generality. 
The unitary evolution under $\hat{U}_{\mathrm{int}}$ generates translation of the position space wave-packet of the pointer and entangles them with the spin  at the $j^{th}$ site as,
\begin{eqnarray}
&|\Psi(\delta t)\rangle& = \hat{U}_{\mathrm{int}} |\psi \rangle \otimes | \phi (x_{c}=0)\rangle \nonumber \\
&=&\Pi_{+}^{(j)} | \psi \rangle \otimes | \phi (x_{c}=\lambda ) \rangle + \Pi_{-}^{(j)}| \psi \rangle \otimes | \phi (x_{c}= -\lambda ) \rangle. \nonumber \\
\end{eqnarray}
Here, $|\Psi(\delta t)\rangle$ denotes the combined state of the system and the pointer after the unitary evolution that consists of two Gaussian states $|\phi(x_c=\pm \lambda )\rangle$ of the same width $\Delta$ with their centers shifted to $x_c=\pm \lambda $ corresponding to the spin eigenstates of $\sigma_{z}^{(j)}$.

The next step is to perform a readout operation on the pointer location by measuring the operator $\hat{M}_C(x_{o}) = \mathbb{I} \otimes |x_{o}\rangle \langle x_{o}|$, where we use the $C$ subscript to denote continuous measurement outcomes $x_0$. 
These set of Krauss operators~\cite{nielsen2010quantum} satisfy the completeness relation: $\int_{-\infty}^{\infty} \hat{M}_C^{\dagger}(x_{o})\hat{M}_C(x_{o}) dx_o = 1$, as required for a POVM.  
The system-pointer state after the readout operation is given by,
\begin{eqnarray}
\label{eq:updatedcts}
|\Psi (\delta t)\rangle &\rightarrow& \frac{\hat{M}_C(x_{o})|\Psi \rangle}{||\hat{M}_C(x_{o})|\Psi \rangle||} 
\\
& = & \frac{1}{\sqrt{p(x_{o};\lambda/\Delta)}} \Big[\Pi_{+}^{(j)} G_{\Delta}(x_{o}-\lambda)  
\nonumber
\\
&+&
\Pi_{-}^{(j)} G_{\Delta}(x_{o}+\lambda ) \Big] | \psi \rangle \otimes \nonumber |x_{o}\rangle.
\nonumber
\end{eqnarray} 
The probability of measuring the pointer at $x_o$ is given by 
\small
\begin{eqnarray}
    p(x_o;\lambda/\Delta) &=& \langle \Psi |\hat{M}_C^{\dagger}(x_{o}) \hat{M}_C(x_{o}) |\Psi \rangle \\
     \! \! \! \! \! &=& \langle \psi|\Pi_{+}^{(j)}| \psi \rangle  G^2_{\Delta}(x_{o}-\lambda ) + \langle \psi|\Pi_{-}^{(j)}| \psi \rangle  G^2_{\Delta}(x_{o}+\lambda ) \nonumber
     .
\end{eqnarray}
\normalsize
$p(x_o)$ is schematically shown in Fig.~\ref{fig:schematicmodel}(b) which consists of two overlapping Gaussians and 
hence, there is no one to one correspondence between the pointer readout position and the spin eigenstates of the qubit. As a result, a readout/projection of pointer location to $x_o$ does not project the state of system to one of eigenstates of $\sigma_z^{(j)}$ and the state of the spin is only weakly measured. 
We recover projective measurements in the limit of non-overlapping Gaussians ($\lambda  \gg \Delta$), as the pointer will only have nonzero probability to be in eigenstates $|x\rangle$ near the peaks of the Gaussian wavepackets, $x_c=\pm \lambda $.

To summarize this subsection, we implement a weak measurement protocol on a qubit by entangling it to a classical pointer whose position can be measured in a continuous basis. For any arbitrary strength of measurements determined by the ratio $\lambda/\Delta$, we perform a readout operation on the pointer with its location $x_o$ sampled from the probability distribution $p(x_{o};\lambda/\Delta)$ and update the system-pointer state governed by Eq.\eqref{eq:updatedcts}. The two step weak-measurement operation is denoted by the operator 
\begin{equation}
    \hat{P}_{\mathrm{CGPM}}=\hat{M}_C(x_o)\hat{U}_{\mathrm{int}}(\delta t).
\end{equation}
In the continuous model the probability density $p(x_o;\lambda/\Delta)$ has the dimension of inverse length and is vanishingly small as it has infinitesimal support in real space. That is, the Born probability has a continuous set of outcomes and within an infinitesimal spatial interval $dx$ goes like 
\begin{equation}
    p_{x}^{\mathrm{CGPM}} = p(x;\lambda/\Delta) d x.
\end{equation}

As a result of the infinitesimal and continuous Born probability, it is not currently straightforward to define the free energy of the log-CFT  based on the Born probabilities using this measurement model (ellaborated on in Sec.~\ref{sec:TM}). It is however, natural to do so in the limit of a discrete number of outcomes with a finite Born probability, and we now therefore turn to constructing a discrete measurement model based on the continuous pointer outcomes.

\subsubsection{Discrete Gaussian pointer } \label{measureDiscrete}
We now consider a weak measurement model where the classical pointer can be measured in discrete positions, $|x_{i}\rangle$ separated by a distance $\epsilon$. We call this the discrete Gaussian pointer model (DGPM). The DGPM can be deduced from the continous measurement model, which replaces the basis states $|x\rangle$ by $|x_{i}\rangle$. The basis states $|x_{i}\rangle$ are obtained by binning the continuous pointer positions $|x\rangle$ over a width of $\epsilon$. The measurement region $\epsilon$ is shown schematically by the blue rectangles in Figure \ref{fig:schematicmodel}(b). This binning gives the orthonormal basis states of the pointer,
\begin{equation}
\vert x_{i} \rangle = \frac{1}{\sqrt{\epsilon}} \int \limits_{x_{i}-\frac{\epsilon}{2}}^{x_{i} + \frac{\epsilon}{2}} \vert x \rangle dx,~ \forall ~i \in \mathbb Z.
\label{eq:discretebasis}
\end{equation}
The pointer is initialized in a Gaussian state ${\vert \phi (x_c=0)\rangle}$ of width $\Delta$ centered at $x_c=0$, which can be expanded in the discrete position basis as, 
\begin{equation}
\vert \phi (x_c=0)\rangle = \frac{1}{\sqrt{\Delta}}\sum \limits_{i=-\infty}^{i=\infty} a_i \vert{x_{i} }\rangle,
\end{equation}
where the corresponding probabilities, $|a_i|^2$s are chosen from a squared Gaussian distribution $G_{\Delta}^2(x)$ averaged over a width $\epsilon$ around $\vert{x_{i}} \rangle$ as,
\begin{equation}
|a_i|^2= \int \limits_{x_{i}-\frac{\epsilon}{2}}^{x_{i} + \frac{\epsilon}{2}}|a(x')|^2dx' = \int \limits_{x_{i}-\frac{\epsilon}{2}}^{x_{i} + \frac{\epsilon}{2}}G_{\Delta}^{2}(x')dx'.
\end{equation}
As previously, the system is coupled to a classical pointer whose Hamiltonian is given by Eq. \eqref{eq:Hint}, but is now measured in a discrete basis.

The readout operation on the pointer location is performed by measuring the operator $\hat{M}_\mathrm{D}(x_{o})=\mathbb{I} \otimes |x_{o}\rangle \langle x_{o}|$ in discrete basis $|x_o \rangle$ given in Eq.\ref{eq:discretebasis} .The measurement operation is now denoted by the operator
\begin{equation}
    \hat{P}_{\mathrm{DGPM}}=\hat{M}_D(x_o)\hat{U}_{\mathrm{int}}(\delta t).
\end{equation}
Thus, the system-pointer state after the readout operation is analogously given by,
\begin{eqnarray}
|\Psi(\delta t)\rangle &\rightarrow &
\frac{\hat{M}_{\mathrm{D}}(x_{o})|\Psi \rangle}{||\hat{M_\mathrm{D}}(x_{o}) | \Psi \rangle ||} 
\nonumber
\\
& =  &\frac{1}{\sqrt{p(x_{o};\lambda/\Delta,\epsilon)}}  \Big[\Pi_{+}^{(j)} \sqrt{p_+(x_{o};\lambda/\Delta,\epsilon)}  \nonumber \\
&+&
\Pi_{-}^{(j)} \sqrt{p_-(x_{o};\lambda/\Delta,\epsilon)} 
\Big ] | \psi \rangle \otimes |x_{o}\rangle,
\label{eq:updatediscrete}
\end{eqnarray} 
where $x_o$ is discrete and $p_\pm(x_{o};\lambda/\Delta,\epsilon)$ denotes the
nonzero probability amplitudes from the  two Gaussian wave-packets shifted to $x_c=\pm \lambda $ that is given by,
\begin{equation}
    p_{\pm}(x_{o};\lambda/\Delta,\epsilon)= \int \limits_{x_{o}-\frac{\epsilon}{2}}^{x_{o} + \frac{\epsilon}{2}}G_{\Delta}^{2}(x'\mp \lambda  )dx' .
    \label{eq:diskrauss}
\end{equation}
Fig. \ref{fig:schematicmodel}(b) illustrates the probability distribution of the classical pointer when two Gaussians significantly overlap. In this case the measured spin will not collapse to an eigenstate of $\sigma_{z}$ (unless it was already in an eigenstate prior to measurement due to the overlap).

\begin{figure*}[t!]\centering
\centering
\includegraphics[scale=0.8]{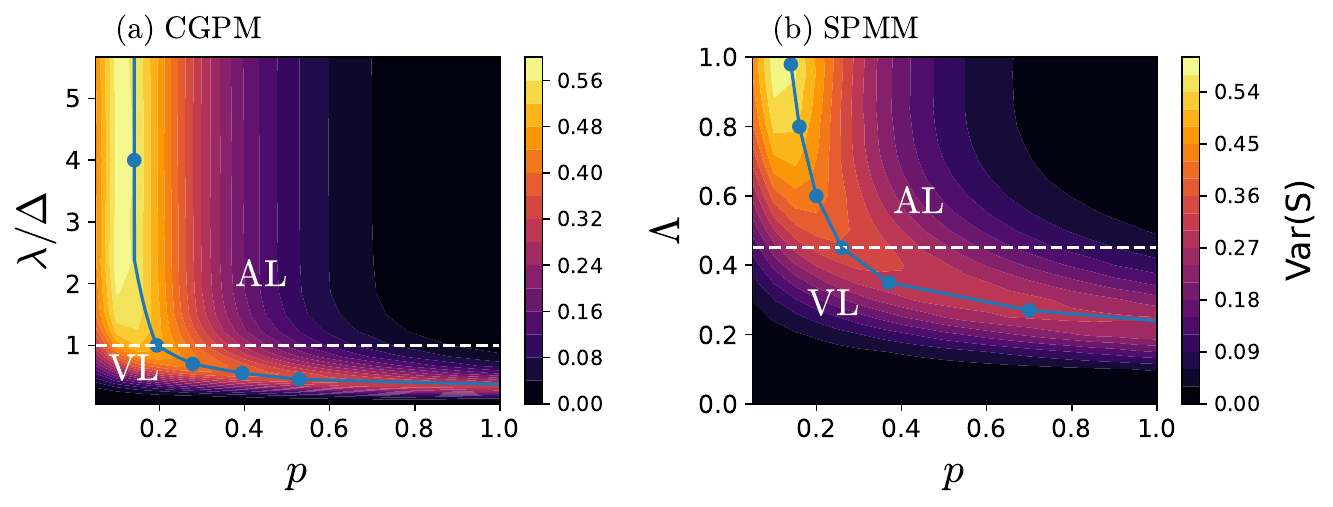}
\caption{
{\bf Phase diagrams for the weak measurement models considered:} 
To qualitatively see the phase boundary we present the variance of the bipartite entanglement entropy, $\mathrm{var}(S)$ as color,  for a chain of $L=12$ qubits with varying measurement rate $p$ and varying strength of measurement $J=\lambda/\Delta$ [introduced in Eq.~\eqref{eq:Hint} for the CGPM defined in Sec.~\ref{measureCont}]  in (a) and $J=\Lambda$ [introduced in Eq.~\eqref{eqn:softprojector} for  the SPMM defined in Sec.~\ref{measureSoft}] in (b). 
The blue circles track the critical measurement rate $p_c(J)$ obtained from the finite-size scaling collapse of the ancilla entanglement entropy $S_{\rm{anc}}$ following Eq.~\eqref{eq:scalecollapse} (the blue line extending to $p=1$ is a guide to the eye). 
For small $p$ and weak coupling strength the models are in the volume law (VL), entangled phase, whereas for large coupling strength and measurement rate the models have long time stead states that are area-law (AL) entangled. In both models, at the phase boundary between the area and volume law phases, the fluctuations in $S$ are effectively maximized shown by the light yellow color. The white line denotes the transitions that  we focus on by computing their critical exponents in Sec.~\ref{sec:phasediagram} and  their log-CFT properties in Sec.~\ref{sec:TM}.   
}
\label{fig:phaseboundary}
\end{figure*}
Now, the Born probability  of measuring the pointer at $x_o$ over width $\epsilon$ is,
\begin{eqnarray}
    p(x_{o};\lambda/\Delta,\epsilon)&=&\langle \psi|\Pi_{+}^{(j)}| \psi \rangle p_+(x_{o};\lambda/\Delta,\epsilon) 
    \nonumber
    \\
    &+& \langle \psi|\Pi_{-}^{(j)}| \psi \rangle p_-(x_{o};\lambda/\Delta,\epsilon).
    \label{eq:pxoDiscrete-main}
\end{eqnarray} 
In the limit of small $\epsilon/\Delta$, we find analytical expressions for the probabilities $p(x_{o};\lambda/\Delta,\epsilon)$ given in Appendix~\ref{sec:AppendixC}. In our numerical calculation, we use at most $\epsilon/\Delta=10^{-5}$ and thus use the Born probabilities given in Eq.\ref{DGPM_analytical}.
In the discrete measurement model, we now have a well-defined Born probability, $p(x_{o};\lambda/\Delta,\epsilon)$ of the measurement outcome measuring the pointer within a bin centered at $x_{o}$.
Therefore, we use the discrete model in computing quantities involving the probability of the measurement record (see Sec.~\ref{sec:TM}). However, we find that the average entanglement properties such as the half-cut and ancilla entanglement entropies are numerically equivalent between the discrete and continuous outcome models as shown in Appendix~\ref{sec:AppendixB}.

In the next subsection, we will discuss a different weak measurement model which does not involve a system-pointer coupling to implement the measurement protocol. 

\subsubsection{Softened Projective Measurement Model}\label{measureSoft}

\begin{figure*}[t!]
\includegraphics[scale=0.75]{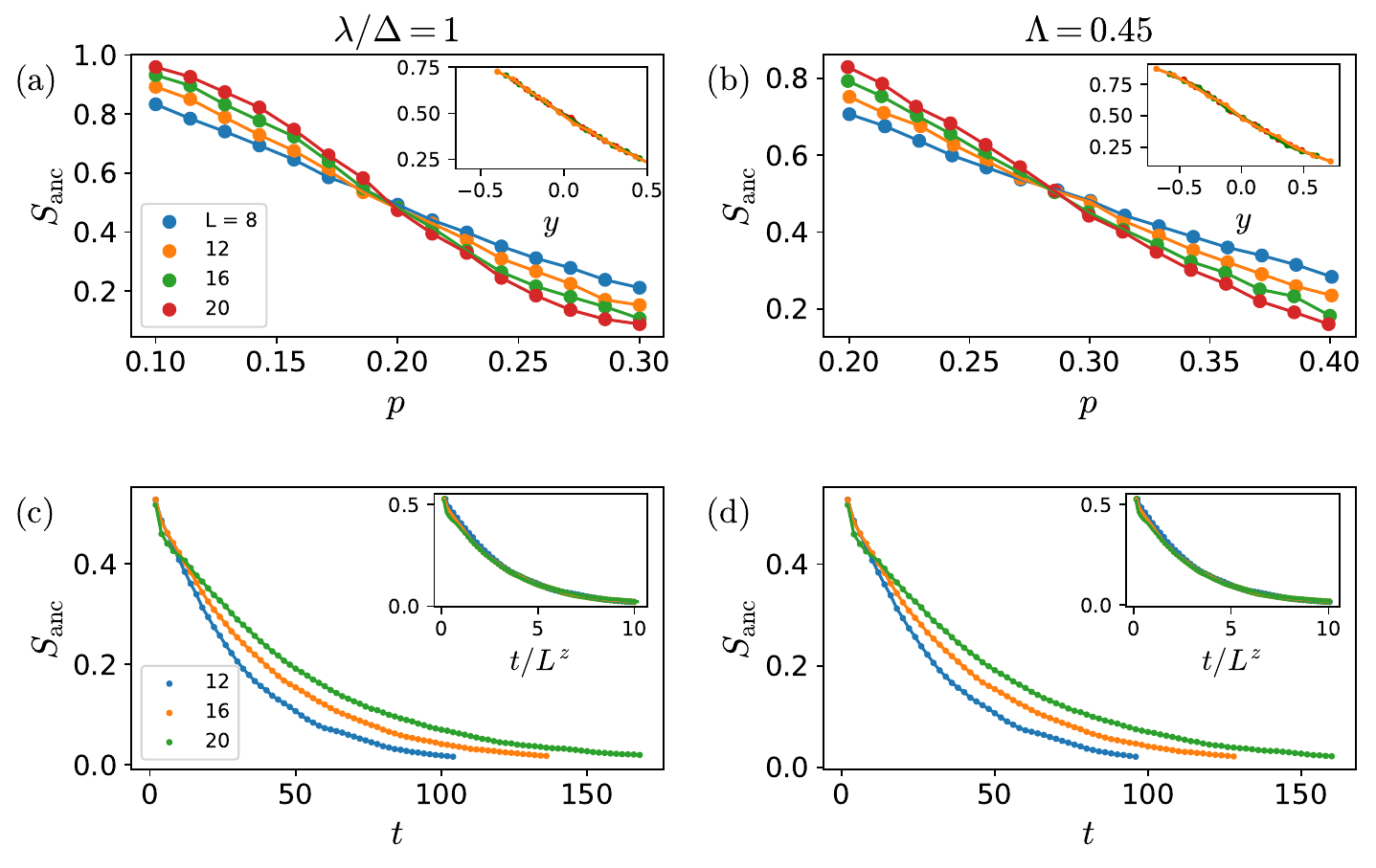}
\caption{{\bf Properties of the ancilla entanglement entropy  and associated critical exponents}: Late time ancilla entanglement entropy $S_{\mathrm{anc}}(t=2L;p,J)$ vs $p$ for the for the the CGPM with $J=\lambda/\Delta=1$ in (a) and the SPMM with $J=\Lambda=0.45$ in (b) show a crossing at the critical point~\cite{GullansHuse-2020,Zabalo-2020} $p=p_c(J)$ in each model. We define $y=(p-p_c)L^{1/\nu}$. The location of MIPT $p_c(J)$ and the correlation length exponent $\nu(J)$ are obtained from a finite size scaling collapse using Eq.~\eqref{eq:scalecollapse} shown in the insets. In both models, we perform the collapse with $\mathrm{L} \in \{12,16,20\}$. We find $p_{c} = 0.19(1)$ and $\nu = 1.3(3)$ for the CGPM with $J=\lambda/\Delta=1$ and $p_{c} = 0.28(2)$ and $\nu = 1.6(3)$ for the SPMM with $J=\Lambda=0.45$. $\nu(J)$ agrees within error-bars with the strongly projective case having $\nu=1.3(3)$.  We comment however that $\nu$ within these proxies is not expected to be sufficiently accurate to make conclusions about the critical properties.
The dynamical critical exponent $z(J)$ is obtained from the time-dependence of $S_{\mathrm{anc}}(t;p_c,J)$ at $p=p_c(J)$ shown in (c) for the CGPM and (d) for the SPMM. Following Eq.~\eqref{eq:scalecollapse}, we collapse $S_{\mathrm{anc}}(t;p_c,J)$ vs $t/L^z$ as shown in the insets and obtain $z(\lambda/\Delta=1) = 0.94(6)$ for the CGPM and $z(\Lambda=0.45) = 0.95(5)$ for the SPMM. In both models, we perform the collapse with $L \in \{12,16,20\}$. 
}
\label{fig:orderparameter}
\end{figure*}

We will now illustrate a weak measurement protocol introduced in Ref.\onlinecite{PhysRevB.108.165120} obtained by softening of the projective measurement in the eigenbasis of $\sigma_z$ by a tunable softening parameter $\Lambda$. We call this model the softened projective measurement model (SPMM). This weak measurement operator at site $j$ is defined as,
\begin{equation}
\hat{P}_{\pm}^{(j)} = \frac{1 \pm \Lambda \sigma_{z}^{(j)}}{\sqrt{2(1+\Lambda^{2}})},
\label{eqn:softprojector}
\end{equation} 
which also satisfies the completeness condition $P_{+} + P_{-} = 1$. Here $\Lambda $ can be varied between $[0,1]$ to control the strength of the measurement. 
Fig.~\ref{fig:schematicmodel}(c) schematically shows the effect of the weak measurement in the Bloch sphere for a single qubit. In the limit $\Lambda=0$, $\hat{P}_{\pm}$ reduces to the identity (up to a multiplicative constant). At $\Lambda = 1$ the weak measurement operators $\{\hat{P}_{+}^{(j)} , \hat{P}_{-}^{(j)} \}$ map onto the strong projector operators $\{\Pi_{j,+}^{z},\Pi_{j,-}^{z}\}$ defined in Eq.~\eqref{eqn:spinproject} implementing measurement along $|\pm z \rangle$ axis. For the intermediate values of $\Lambda$, $\hat{P}_{\pm}$ interpolates between  $|\pm z\rangle$ and $|\pm n\rangle$. 
To perform the weak measurement, we calculate the Born probabilities $p_{\pm}(\Lambda)$ corresponding to the operators $\hat{P}_{\pm}^{(j)}$ from the system wave-function $|\psi \rangle$ as,
\begin{eqnarray}
p_{\pm}(\Lambda)&& = \langle \psi |\hat{P}_{\pm}^{(j)}\hat{P}_{\pm}^{(j)} | \psi \rangle \nonumber \\
&&= \frac{1}{2(1+\Lambda^{2})}(1 + \Lambda^{2} \pm 2 \Lambda\langle \psi | \sigma_{z}^{(j)} | \psi \rangle). 
\label{eq:psoft}
\end{eqnarray}
Based on $p_{\pm}(\Lambda)$, we measure either $\hat{P}_{+}^{(j)}$ or $\hat{P}_{-}^{(j)}$ and update the state of the system to 
\begin{equation}
    | \psi \rangle \rightarrow \frac{\hat{P}_{\pm}^{(j)}|\psi \rangle}{||\hat{P}_{\pm}^{(j)}|\psi \rangle||}.
\end{equation}
This implements a soften version of the standard projective measurement controlled by the parameter $\Lambda$.

\section{Phase Diagrams of  weak MIPTs}\label{sec:phasediagram}
In this section, we will locate the entanglement phase transition as a function of measurement strength ($J$) and rate ($p$). To get a qualitative sense of the phase diagram, we use the variance of the bipartite entanglement entropy as a function of measurement strength and probability in Sec.~\ref{probhalfcut}. To provide an accurate, unbiased estimate of the location of the MIPT we use finite size scaling of an ancilla based local order parameter~\cite{GullansHuse-2020,Zabalo-2020} in Sec.~\ref{orderparameter}.  
To compute all the Reyni entropies (e.g. the half-cut bipartite Renyi entropy and the ancilla Renyi entropies defined below)  in the models with a classical pointer coupled to the system, we use the CGPM.
As we show in Appendix~\ref{sec:AppendixB}, both the CGPM and DGPM yield the same numerical value for the Reyni entropies, up to negligible numerical fluctuations of the order $10^{-3}$ times smaller than the value of the Reyni entropies. 

\begin{figure*}[t!]
\includegraphics[scale = 0.75]{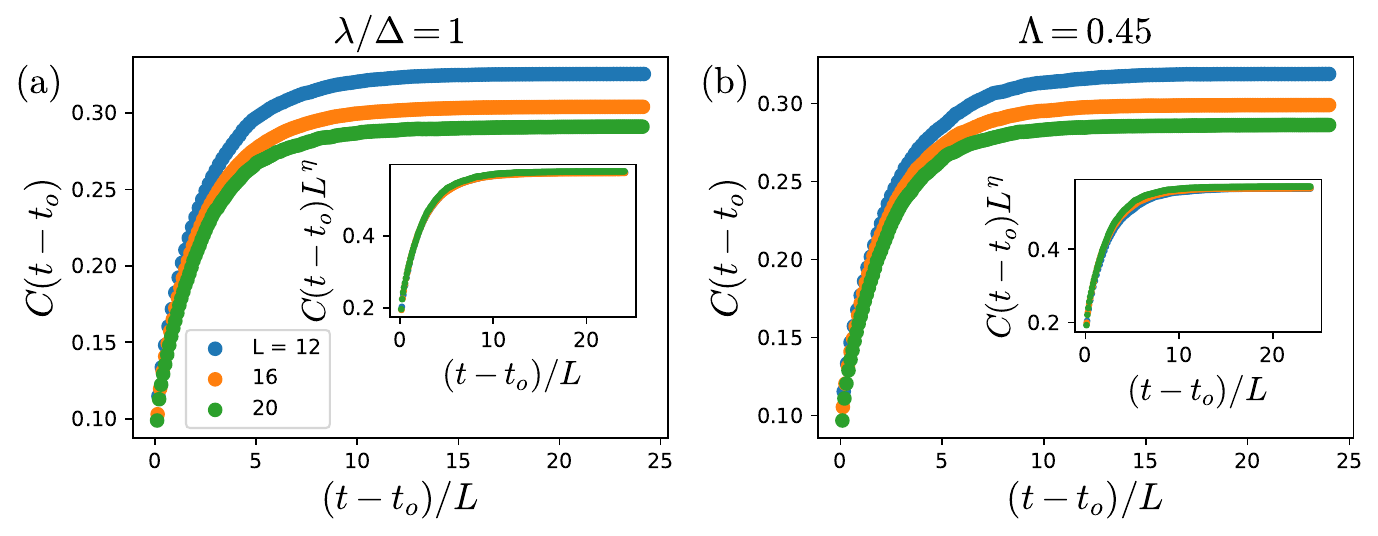}
\caption{{\bf Mutual information between two ancillas at the critical point and exponent $\eta$:} (a) Mutual information between two ancillas entangled at $r=1$ and $r'=\frac{L}{2}$ 
in the steady state of the circuit evolving with the CGPM at $\lambda/\Delta= 1$ and $p=p_{c} = 0.19(3)$ shown in (a) and that for SPMM at $\Lambda= 0.45$ and $p=p_{c} = 0.28(2)$ shown in (b). 
We use three different system sizes $L=12, 16, 20$ for the scaling collapse following Eq.~\eqref{eq:corrlcollapse} shown in the inset. We find the bulk exponent $\eta = 0.19(3)$ in SPMM and $\eta = 0.21(2)$ in CGPM. We average over $10^4$ trajectories for each system size.
}
\label{fig:corrlCGPM}
\end{figure*}

\subsection{Bipartite entanglement entropy}\label{probhalfcut}
We begin by estimating the phase boundary from fluctuations in the bipartite entanglement entropy. To compute the bipartite entanglement entropy, we divide the system into two halves (denoted A and B) and the reduced density matrix $\rho_A =\rm{Tr}_B[|\psi(t)\rangle\langle \psi(t)|]$ is used to compute the half-cut von Neumann entanglement entropy
\begin{equation}
    S_{n=1}= -\rm{Tr}_A[\rho_A \ln \rho_A].
\end{equation}
Following Ref.~\onlinecite{WeakSchomerus}, we compute the variance of the bipartite entanglement entropy $\mathrm{var}(S)$ as a qualitative proxy of the phase diagram.
We compute the variance from an ensemble of entanglement entropy values constituted from (i) the steady state $S(t=2L)$ values from different quantum trajectories~\cite{Aidan_PRL} and (ii) $S(t)$ values from a single trajectory choosing $t$ from a quasi-stationary regime, $t=L/2$ to $100$. This steady state regime 
is chosen such that the initial growth of the entropy with time has saturated. Fig.~\ref{fig:phaseboundary} shows $\rm{var}(S)$ in the two-dimensional parameter space spanned by $p$ and $J$ with a color plot where the lighter color indicates larger variance. Fig.~\ref{fig:phaseboundary}(a) corresponds to the CGPM  while Fig.~\ref{fig:phaseboundary}(b) is for the SPMM. We average the bipartite entanglement entropy over 3000 trajectories for CGPM and 2000 trajectories for SPMM for each values of $p$ and $J$.
The variance is expected to be maximal at the critical measurement rate $p=p_c(J)$ (where  $J=\lambda/\Delta$ for CGPM/DGPM and $J=\Lambda$ for SPMM), and as shown in the data we see a clear maximum in the parameter space for a fixed system size $L=12$. However, the location of this maximum is known to drift with increasing system size~\cite{WeakSchomerus}  and therefore in the next section we consider a separate unbiased estimate of $p_c(J)$. In particular, we use the ancilla entanglement entropy defined below in the next subsection, and its crossing for various system sizes as a function of $p$ to locate $p_c(J)$.

\subsection{Ancilla Qubit Order Parameter}\label{orderparameter}
To construct a local order parameter of the MIPT  \cite{GullansHuse-2020}, we couple an ancilla qubit locally by putting it in a Bell pair  with a spin in the system. 
We then apply an encoding step to scramble the locally entangled ancilla by running the circuit without measurements and only unitary gates out to $t_{o}=2L$, which  prepares the ancilla in a maximally entangled states with the system. We then run the hybrid measurement and unitary dynamics and call this time $t=0$ in the results presented.

As a function of time we calculate the ancilla von Neumann entanglement entropy $S_{\rm{anc}}$ of the reduced density matrix of the ancilla after integrating out all of the spins in the circuit. 
In the monitored dynamics, $S_{\rm{anc}}(t;p,J)$ decreases monotonically with time from its maximum value $1$ as time increases. 
The ancilla entanglement entropy $S_{\rm{anc}}(t;p,J)$ serves as an order parameter for the MIPT in the steady state ($t\approx 2L$)~\cite{GullansHuse-2020,Zabalo-2020}.  To estimate the critical point $p_c(J)$, we perform a finite-size scaling with the following ansatz 
\begin{equation}
S_{\rm{anc}}(t;p,L) \sim Q((p-p_{c})L^{1/\nu},t/L^{z}),
\label{eq:scalecollapse}
\end{equation}
where $Q(x,y)$ is an arbitrary scaling function, $\nu(J)$ and $z(J)$ are the correlation length exponent and the dynamical exponents of the MIPT, respectively.

We show the data for $S_{\rm{anc}}(t= 2L;p,J)$ vs $p$ for the for the 
CGPM model with $J=\lambda/\Delta=1$ in Fig.~\ref{fig:orderparameter}(a) and SPMM model with $J=\Lambda=0.45$ in Fig.~\ref{fig:orderparameter}(b), in which both display a clear crossing at $p_c(J)$ for various systems sizes, consistent with the scaling ansatz in Eq.~\eqref{eq:scalecollapse}. Fixing the aspect ratio of time and the system-size ($t=2L$), we collapse the ancilla order parameter for different system sizes $L=12,16,20$ to estimate $p_c(J)$ and $\nu(J)$ as shown in the insets of Fig.~\ref{fig:orderparameter}(a) and (b). This yields $p_c(\lambda/\Delta=1)=0.19(1)$ and $\nu(\lambda/\Delta=1)= 1.3(3)$ for the CGPM and $p_c(\Lambda=0.45)=0.28(2)$ and $\nu(\Lambda=0.45)= 1.6(3)$ for the SPMM. These estimates of $\nu$  agree within their numerical accuracy with the value for the projective measurement case $\nu \approx 1.2(3)$ \cite{Zabalo-2020}. We average $S_{\rm{anc}}(t= 2L;p,J)$ over 3,000 trajectories for CGPM and 2,000 trajectories in SPMM for each values of $p,L$ and $J$.

Varying the strength of the measurements $J$, we calculate
$p_c(\lambda/\Delta)$ for the CGPM and $p_c(\Lambda)$ for the SPMM shown in Fig.\ref{fig:phaseboundary} (a) and (b) respectively as solid blue circles. The critical transition rate $p_c(J)$ increases with decreasing strength of measurement $J(=\lambda/\Delta~\mathrm{ or}~ \Lambda)$ in both the models as expected. This estimation of $p_c$ from the ancilla order parameter ($S_{\rm{anc}}$) clearly follows (though not precisely) the locus of the maximum of  var$(S)$ (at fixed system size $L=12$) that is shown by bright yellow color. 
Interestingly, in both models we find that if the measurement strength is too weak its not possible to drive a phase transition out of the volume law phase with measurements. 

To estimate the dynamical exponent $z$, we study the time-dependence of the ancilla order-parameter at the critical measurement $p=p_c$. In Fig.~\ref{fig:orderparameter} 
we show $S_{\rm{anc}}(t;p_c,J)$ as a function of $t$ for the CPGM with $\lambda/\Delta=1$ in (c) and for the SPMM with $\Lambda=0.45$ in (d) for three different system-sizes $L=12,16,20$. We average the time-dependent $S_{\rm{anc}}(t;p_c,J)$ over 10,000 trajectories for both CGPM and and SPMM for each values of $L$ and $J$.
We perform a finite size scaling collapse of $S_{\rm{anc}}(t;p_c,J)$ as a function of $t/L^{z(J)}$ using Eq.\ref{eq:scalecollapse} and the collapsed curves are shown in the insets of Fig.~\ref{fig:orderparameter} (c) and (d). We find $z(\lambda/\Delta = 1) = 0.94(6)$ and $z(\Lambda = 0.45) = 0.95(5)$. Thus our results provide strong numerical evidence of conformal invariance at the weak measurement induced transition.

\subsection{Mutual information between two ancilla qubits}

We next compute the anomalous dimension exponent $\eta(J)$ following the protocol prescribed in Ref.\onlinecite{GullansHuse-2020,Zabalo-2020}. 
We initialize the system in a random product state and then evolve the circuit with monitored dynamics to reach a steady state at time $t_{o} = 20L$. 
We note that in HDU circuit this wait time is much longer to reach the steady  state for the mutual information between two ancillas compared to that with Haar random gates (where $t_o\approx 2L$), though we find this is important to reach a stable (i.e. wait time independent) estimate of $\eta$.
We introduce two ancilla qubits $\tilde{A}$ and $\tilde{B}$, and maximally entangle them with the circuit at the spacetime points $(r,t_{o})$ and $(r',t_{o})$. We define the connected order-parameter correlation function $C(t-t_{o})$ as the mutual information between the ancillas $\tilde{A}$ and $\tilde{B}$. In both the volume law phase, $p<p_{c}$, and area law phase, $p>p_{c}$, $C(t-t_{o}) \sim \mathrm{exp}(-L/\xi)$ for $L \gg \xi$ where $\xi\sim (p-p_{c})^{\nu}$ is the finite correlation length at generic $p \neq p_{c}$. At MIPT $p=p_c(J)$, $C(t-t_o)\sim 1/L^{\eta}$. 
To compute the bulk exponent $\eta$, we entangle the two ancillas to antipodal sites $(r-r' = L/2)$ with periodic boundary conditions in the circuit and perform a finite-size scaling with the following ansatz (assuming $z=1$) 
\begin{equation}
C(t-t_{o}) \sim  L^{-\eta} \tilde Q((t-t_o)/L).
\label{eq:corrlcollapse}
\end{equation}
where  $\tilde Q(x)$ is an arbitrary scaling function. In
Fig.\ref{fig:corrlCGPM}(a) and (b) we show $C(t-t_o)$ vs $t-t_o$ for the CGPM model at $\lambda/\Delta=1$ and $p=p_c= 0.19(1)$ and the SPMM model at $\Lambda=0.45$ and $p=p_c=0.28(2)$ for three different system sizes $L=12,16,20$. In the data presented, we average over $10^4$ trajectories for each system size.
The scaling collapse is shown in the inset yielding $\eta = 0.21(3)$ for CGPM and $\eta=0.19(2)$ for the SPMM. We also compute the bulk exponent $\eta$ for the projective measurement case in the HDU circuit in Appendix~\ref{sec:HDU} and find $\eta = 0.23(2)$. Hence, based on numerical estimation of the exponents $\eta$ and $\nu$, we observe that weakening the strength of measurements does not change the bulk exponents of the MIPT.

\section{Transfer Matrix Approach and Properties of the log-CFT}
\label{sec:TM}
At the critical point $p=p_{c}$ of the MIPT we have shown numerical evidence of Lorentz invariance at the transition in Figs.~\ref{fig:orderparameter} and \ref{fig:corrlCGPM}. Through mappings to classical statistical mechanics models via an infinite onsite Hilbert space dimension~\cite{PhysRevB.100.134203,JianMIPT,AltmanPRB} or the zeroth Renyi entropy~\cite{NahumPRX} a firm connection to percolation in $1+1$-dimensions, which is a well known example of a log-CFT~\cite{LUDWIG1987687,Cardy_2013,GURARIE_LUDWIG2005,GURARIE1993LogCFTs}, has been established. Moving away from these tractable limits numerical evidence for the nature of the log-CFT in qubit chains with strong projective measurements was unveiled through a transfer matrix description that probe the bulk~\cite{Aidan_PRL} and the boundary~\cite{kumar2023boundary} critical exponents. Focusing on the bulk Lyapunov spectrum of this non-unitary transfer matrix, the effective central charge of the log-CFT, the leading typical scaling dimension of the order parameter, and the potential multifractal nature of the transition  can be computed numerically by utilizing  numerical techniques from percolation applied to the MIPT~\cite{Cardy_doublefitting,jacobsen1998critical,Aidan_PRL}. In the following section, motivated by this past work and the numerically observed Lorenz invariance of the transition in Sec.~\ref{sec:phasediagram}, we utilize this transfer matrix description to study the nature of the  log-CFT governing the MIPT with weak measurements.

\subsection{Lyapunov Spectrum of the Transfer Matrix} 
% Define m
We adopt the transfer matrix method introduced in Ref.\onlinecite{Aidan_PRL} to describe the non-unitary evolution of the hybrid circuit. We summarise the key ingredients of the method here for both the sake of completeness and to understand the reason why we construct the DGPM. The time-evolution of a quantum circuit is represented as an ensemble of quantum trajectories where each trajectory is defined by a fixed set of unitary gates and the location and time of measurement operations. Each time-step of a trajectory is defined by a Krauss operator $K_t^{\vec{m}}=P_t^{\vec{m}}U_t$ where $U_t$ are the unitary gates acting at each time step and $P_t^{\vec{m}}$ is the weak measurement operator defined in Sec.~\ref{sec:model}. The Krauss operators at time $t$ depends on the history of measurement outcomes $\vec{m}$ in that trajectory. 
In the DGPM $\vec{m}=(x_{o}^1,x_{o}^2,x_{o}^3,\dots)$ where $x_{o}^i$ refers to the measured pointer position at the $ith$ measurement event and take an infinite number of possibilities. Whereas, for the SPMM $\vec{m}=(m^1,m^2,m^3,\dots)$ and $m^i=\pm 1$ for the two possible outcomes given in Eq.~\eqref{eq:psoft}.
The time evolution of the density matrix in a trajectory is represented as,
\begin{equation}
    \rho(t)=\frac{1}{Z_{\vec{m}}}K_{\vec{m}} \rho(t=0) K^{\dagger}_{\vec{m}},
\end{equation}
where $K_{\vec{m}}=\prod_{t'=0}^{t} K_{t'}^{\vec{m}} $ and $Z_{\vec{m}}$ is the partition function of the statistical mechanics model describing the trajectory~\cite{JianMIPT,Aidan_PRL} and is given by,
\begin{equation}
    Z_{\vec{m}}(t)= \mathrm{Tr}\left [ K_{\vec{m}} \rho(t=0) K^{\dagger}_{\vec{m}}\right] = \sum_i e^{\lambda_i^{\vec{m}}t}. 
\end{equation}
Here $\lambda_i^{\vec{m}}$ are the Lyapunov exponents governing the exponential decays in the partition function in the long time limit ($\lambda_{i}^{\vec{m}} < 0)$. The average Lyapunov exponents $\lambda_{0}, \lambda_{1}, ...$ are obtained by averaging $\lambda_{i}^{\vec{m}}$ over trajectories $\vec{m}$ weighing by their corresponding Born probability $p_{\vec{m}}$, i.e. $\lambda_{i} = \sum_{\vec{m}} p_{\vec{m}} \lambda_{i}^{\vec{m}}$.
In the next subsections, we will study the average first two leading Lyapunov exponents $\lambda_0$ and $\lambda_1$ to extract critical exponents governing the log-CFT.

\subsection{Free Energy}
The leading Lyapunov exponent is related to the free energy $F = - \lambda_{0}t$ which can be calculated
as the Shannon entropy of the measurement record $\vec{m}$ averaging over quantum trajectories,
\begin{equation}
F = -\sum_{\vec{m}} p_{\vec{m}}\ln p_{\vec{m}} =- \langle \ln p_{\vec{m}} \rangle.
\label{eqn:freeenergy}
\end{equation}
Here, we point out that a discrete measurement outcome is required for the average of $\ln p_{\vec{m}}$ to be well defined.
At the critical point of the MIPT $p=p_c$, the conformal invariance of the system dictates that the free energy density $f(L,t)=F(L,t)/A$ where the (implicitly defined) space-time area is given by $A=\alpha Lt$. The space time anisotropy factor $\alpha_{\mathrm{HDU}}=1$ for the HDU gates we consider here~\cite{Aidan_PRL}, and this exact knowledge of $\alpha$ is why we are considering these HDU gates.
If we consider Haar unitary gates on the other hand, a direct computation of $\alpha$ through mutual information in space and time~\cite{Aidan_PRL}, yields $\alpha_{\mathrm{HU}}=0.81(9)$ that introduces an additional error in the free energy calculation.
The free energy density after long times ($t\gg L$) for spatial periodic boundary conditions decays with the system size as,
\begin{equation}
    f(L) = f(L=\infty) - \frac{\pi c_{\rm{eff}}}{6 L^{2}} + O(L^{3}),
    \label{eqn:freeEdensity}
\end{equation}
\label{PBCfscaling}
where $c_{\mathrm{eff}}$ is the effective central charge of the log-CFT.

\begin{figure*}[t!]
   \centering
\includegraphics[scale=0.36]{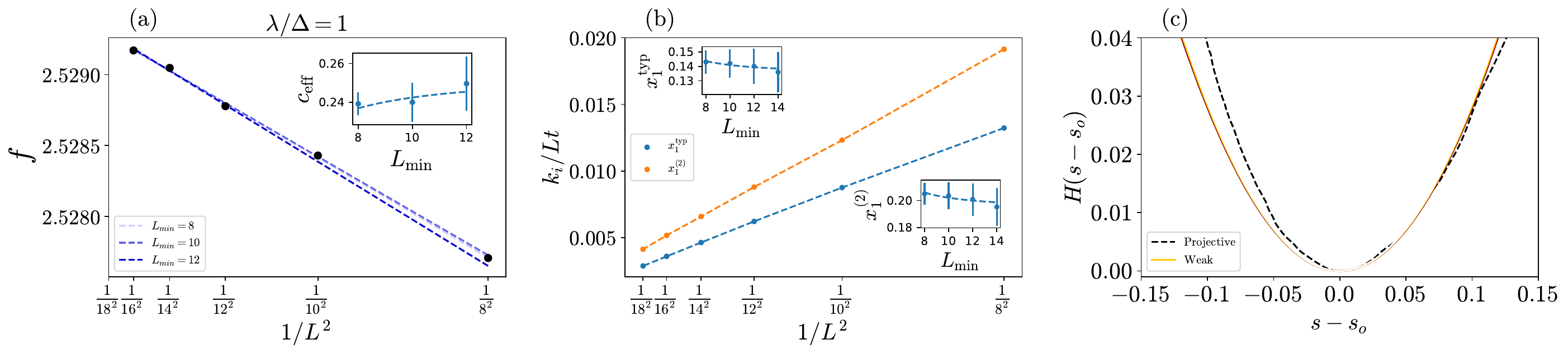}
\caption{{\bf Properties of log-CFT in DGPM model:} (a) shows the decay of the free energy density $f(L)$ defined in Eqs.~\eqref{eqn:freeEdensity} and~\eqref{eqn:freeenergy} vs $1/L^2$ by the black circles with its slope related to the effective central charge, $c_{\mathrm{eff}}$. We perform the double fitting procedure from \cite{Aidan_PRL} by successively removing the smallest system size from the fit, and then fitting the data with $L \geq L_{\mathrm{min}}$, for $L_{\mathrm{min}} \in$ \{8,10,12\} shown by the dashed lines where darker color denotes larger $L_{\mathrm{min}}$. The inset shows $c_{\mathrm{eff}}(L_{\mathrm{min}})$ and we use $c_{\mathrm{eff}}(L_{\mathrm{min}})= c_{\mathrm{eff}}(L_{\mathrm{min}}=\infty)+ b/L^2_{\mathrm{min}}$ to obtain $c_{\mathrm{eff}}(L_{\mathrm{min}}=\infty)=0.25(3)$ quoted in the main text. (b) shows the decay of $k_i/Lt$ vs $1/L^2$ where $k_1$ and $k_2$ are the first two cumulants of the correlation functions ($C^{\vec{m}}$) defined in Eq.~\eqref{eq:cumulants}. The slope of $k_1/Lt$ gives the typical anomalous scaling dimension $x_1^{\mathrm{typ}}=0.14(2)$ while that of $k_2/Lt$ gives the leading multifractal exponent $x_1^{(2)}=0.19(2)$. The presence of multifractality is further confirmed in (c) by showing the scaling collapse of the distribution of $Y(t)=-\ln C^{\vec{m}}(t)$ onto a universal scaling function $H(s)$, following a multifractal scaling given in Eq.~\eqref{eq:multifracscaling}. The values are shown for different system sizes from $L=8$ to $L=18$, and using data from $t=5L$ to $t=32L$, in the solid yellow lines. The darker color denotes a larger system size. The black dashed line shows the $H(s)$ curve extracted from Ref.~\onlinecite{Aidan_PRL}. The x-axis is plotted as $s-s_{o}$, where $H(s)$ attains a minimum at $s_{o} = 0.14 \approx x_1^{\mathrm{typ}}$. We use $\epsilon = 0.1$, $\lambda = \Delta = 10^4$, $p=p_{c}= 0.19$ and periodic boundary conditions for all the plots. We use $3\times 10^6$ quantum trajectories for statistical averaging of the $f$ and $1.5 \times 10^5$ quantum trajectories for averaging of the cumulants. 
}
\label{fig:LyapunovDGPM}
\end{figure*} 

We first calculate the free energy for the weak measurement model with 
the DGPM
illustrated in Sec.\ref{measureDiscrete}. In this model, the Born probabilities of the measurement record are calculated from,
\begin{equation}
    p_{\vec{m}}= \prod_i p(x_o^{i};\lambda/\Delta,\epsilon),
\end{equation}
where $p(x^i_o;\lambda/\Delta,\epsilon)$ given in Eq.~\eqref{eq:pxoDiscrete-main} [and in the limit $\epsilon \ll \Delta$ by   Eq.~\eqref{eq:pxoDiscrete}] is the probability of measuring the classical pointer in a region of width $\epsilon$ around $x_o$ in the $i$th measurement operation in the trajectory $\vec{m}$. We note that $f(L)$ in this case explicitly depends on the choice of $\epsilon$ which is the smallest length scale we keep in our calculation to avoid the vanishingly small probabilities in the continuous measurement case (see Sec.\ref{measureCont} for the limiting case $\epsilon \rightarrow 0$). However, in the limit $\epsilon \ll \Delta$, 
a change in $\epsilon$ leads to a change in the free energy density as $\Delta f = p_c \ln \Delta \epsilon$ leaving our estimate of $c_{\mathrm{eff}}$ unaffected in this limit (see Appendix~\ref{app:ceff} for more details). On the other hand in the strong projective limit, $\lambda/\Delta \rightarrow \infty$, the probabilities $p(x^i_o;\lambda/\Delta,\epsilon)$ reduce to  measuring the spin in the up or down state up to a multiplicative constant. As we show in detail in Appendix~\ref{app:ceff}, the effective central charge also remains invariant w.r.t. changes in $\epsilon$ in the projective measurement limit.
We compute the free energy $F(L,t)$ starting from an initial product state of the circuit. To eliminate the effects of initial conditions in the free energy calculated from the cumulative Born probabilities $p_{\vec{m}}(t)$, we waited till $t=5L$ before starting to record $p_{\vec{m}}(t)$. In the long time limit ($t> 5L$), $F(L,t)$ grows linearly with time and we extract $f(L)$ from its slope between $t=5L$ and $t=32L$.
Fig.\ref{fig:LyapunovDGPM}(a) shows the free energy density $f(L)$ vs $1/L^2$ for different system sizes $L=8$ to $18$ with black circles at the strength of measurement $\lambda/\Delta = 1$.

The finite size scaling form in Eq.~\eqref{eqn:freeEdensity} provides an essential guide to obtaining the correct numerical  estimate of the free energy. An important point to note is that
the DGPM
requires us to average over a much larger number of trajectories ($\sim 2\times10^6$) compared to the strong projective case ($\sim 2.5\times10^4$) 
to see the  $1/L^2$ behaviour dictated by the log-CFT. To estimate $c_{\rm{eff}}$ from its slope, we use a double fitting procedure systematically eliminating the effects of the smaller systems sizes~\cite{Aidan_PRL,Cardy_doublefitting}. We fit the data from $L=L_{\rm{min}}$ to $L=18$ shown by dashed lines and estimate $c_{\rm{eff}}(L_{\mathrm{min}})$ shown in the inset. Extrapolating $L_{\mathrm{min}} \rightarrow \infty$, we obtain the effective central charge $c_{\mathrm{eff}}=0.25(3)$ in a weakly measured DGPM at $p_c(\lambda=\Delta)=0.19(1)$, which matches quite well with the case of strong projective measurement that finds $c_{\mathrm{eff}}=0.25(3)$ in ~\cite{Aidan_PRL}.  

We next calculate the free energy for the softened projective measurement model illustrated in Sec.~\ref{measureSoft}. In this case the Born probabilities $p_{\vec{m}}$ are calculated from,
\begin{equation}
    p_{\vec{m}} = \prod_{i} p_{m^i}(\Lambda)
\end{equation}
where $p_{m^i}(\Lambda)$ are  Born probabilities of individual measurement event (denoted by $m^i=\pm$) defined in Eq.~\eqref{eq:psoft}.
Fig.~\ref{fig:Lyapunov_soft}(a) shows $f(L)$ vs $1/L^2$ at $\Lambda=0.45$ having $p_c=0.28(2)$. We averaged over $25000$ quantum trajectories initialized in both Haar random and product states and waited till $t=4L$ before starting to record Born probabilities. We extract the slope of $f(L)$ vs $1/L^2$ using the above-explained double fitting procedure to obtain $c_{\mathrm{eff}}(L_{\rm{min}})$ shown in the inset. This gives $c_{\mathrm{eff}}=0.26(2)$ in the softened projective measurement case which also closely matches with the projective measurement case. Thus weakening strength of measurements does not affect the effective central charge of the log-CFT governing the MIPT.

\subsection{Leading Scaling Dimension  }
\begin{figure*}[t!]
   \centering
\includegraphics[scale=0.36]{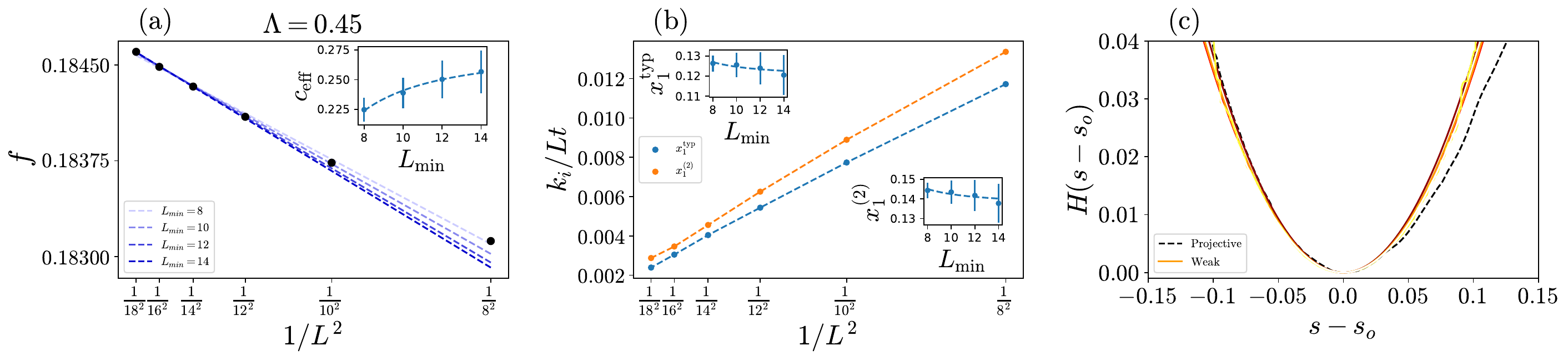}
\caption{{\bf Properties of log-CFT in SPMM model:} % Put the L_{min} values
(a) shows the decay of the free energy density $f(L)$ Eqs.~\eqref{eqn:freeEdensity} and~\eqref{eqn:freeenergy} vs $1/L^2$ by the black circles with its slope related to the effective central charge, $c_{\mathrm{eff}}$. We perform the double fitting procedure\cite{Aidan_PRL} by successively removing the smallest system size from the fit, and then fitting the data with $L \geq L_{\mathrm{min}}$, for $L_{\mathrm{min}} \in$ \{8,10,12,14\} shown by the dashed lines where darker color denotes larger $L_{\mathrm{min}}$. The inset shows $c_{\mathrm{eff}}(L_{\mathrm{min}})$ and we use $c_{\mathrm{eff}}(L_{\mathrm{min}})= c_{\mathrm{eff}}(L_{\mathrm{min}}=\infty)+ b/L^2_{\mathrm{min}}$ to obtain $c_{\mathrm{eff}}(L_{\mathrm{min}}=\infty)=0.26(2)$ quoted in the main text. (b) shows the decay of $k_i/Lt$ vs $1/L^2$ where $k_1$ and $k_2$ are the first two cumulants of the correlation functions ($C^{\vec{m}}$) defined in Eq.~\eqref{eq:cumulants}. The slope of $k_1/Lt$ gives the typical anomalous scaling dimension $x_1^{\mathrm{typ}}=0.12(2)$ while that of $k_2/Lt$ gives the leading multifractal exponent $x_1^{(2)}=0.14(2)$.The presence of multifractality is further confirmed in (c) by showing the scaling collapse of the distribution of $Y(t)=-\ln C^{\vec{m}}(t)$ onto a universal scaling function $H(s)$, following a multifractal scaling given in Eq.~\eqref{eq:multifracscaling}. The values are shown for different system sizes from $L=8$ to $L=18$, using data from $t=5L$ to $t=32L$, in the solid yellow lines. The darker color denotes a larger system size. The black dashed line shows the $H(s)$ curve for the projective measurement case extracted from Ref.~\onlinecite{Aidan_PRL}. The x-axis is plotted as $s-s_{o}$, where $H(s)$ attains a minimum at $s_{o} = 0.12 \approx x_1^{\mathrm{typ}}$. We use  $\Lambda = 0.45$, $p=p_{c}= 0.28$ and periodic boundary conditions for all the plots. We use $2.5\times 10^4$ quantum trajectories for statistical averaging.}
\label{fig:Lyapunov_soft}
\end{figure*} 
In this section we calculate the next leading  Lyapunov exponent $\lambda_1$ and the associated typical and multifractal critical exponents following Ref.~\cite{Aidan_PRL}. To calculate $\lambda_1$, we construct the two orthogonal states of the system $|\psi_1 \rangle$ and $|\psi_2 \rangle$. These states are then evolved in time where at each time step they are subjected to identical Krauss operators $K_{t}^{\vec{m}} $ with the measurement operators, $P_t^{\vec{m}}$s solely determined by the Born probabilities calculated from $|\psi_1 \rangle$. After each  timestep, we orthogonalize $|\psi_{1} \rangle$ to $|\psi_{2} \rangle$ with a Gram-Schmidt projector $P^{GS}_{t}$, which is akin to an additional measurement operation. 
The Born probability of the measurement records in the trajectory $\vec{m}$ corresponding the state $|\psi_2 \rangle$ is written as,
\begin{equation}
p'_{\vec{m}}(t) = ||\Pi_{t'=0}^{t} P^{GS}_{t'} K_{t'}^{\vec{m}} |\psi_{2} \rangle||^{2}.
\end{equation}
This gives the first generalized free energy $F_{1}(L,t) =-\lambda_1 t=- \sum_{\vec{m}} p'_{\vec{m}} \ln p'_{\vec{m}}$. We initialize the system either in Haar state or product state and wait till $t=5L$ before starting to record $p'_{\vec{m}}(t)$s to eliminate the effects of the choice of initial conditions. In the long time limit ($t>5L$), $F_1(L,t)$ is expected to increase linearly in time and we compute the first generalized free energy density, $f_1(L)=F_1(L,t)/(Lt)$ from its slope between $t=5L$ and $t=32L$. 

The log-CFT governing the MIPT suggests that difference between the first two generalized free energy densities decay with $L$ as (taking $\alpha=1$),
\begin{equation}
\frac{k_1}{Lt}=f_{1}(L) - f(L) = \frac{2 \pi x_{1}^{\rm{typ}}}{L^{2}},
\label{eq:k1}
\end{equation}
with a slope related to the typical scaling dimension  of the order parameter, $x_{1}^{\rm{typ}}$. The exponent $x_{1}^{\rm{typ}}$ is related to the bulk exponent $\eta$ (computed in Sec.~\ref{sec:phasediagram} through Eq.~\eqref{eq:corrlcollapse}, and shown in Fig.~\ref{fig:corrlCGPM}) as $x_{1}^{\rm{typ}}=\eta/2$. 

We show the first cumulant
$k_1/Lt$ given in Eq.\eqref{eq:k1} and (taking $\alpha=1$) vs $1/L^2$ in  Fig.\ref{fig:LyapunovDGPM}(b) for the DGPM at $\lambda=\Delta$ and in Fig.\ref{fig:Lyapunov_soft}(b) for the SPMM at $\Lambda=0.45$. 
From the slope of $k_1/Lt$, we find $x_1^{\mathrm{typ}}(\lambda=\Delta)=0.14(2)$ for the DGPM and $x_1^{\mathrm{typ}}(\Lambda=0.45)=0.12(2)$ for the SPMM, which closely agrees with the strong projective case having $x_1^{\mathrm{typ}}=0.122(1)$ obtained in Ref~\onlinecite{Aidan_PRL}. Moreover, our results are consistent with $x_1^{\mathrm{typ}}=\eta/2$ in both the models: In the SPMM, $\eta = 0.19(3)$ and $x_{1}^{\mathrm{typ}} = 0.12(2)$, and in the CGPM/DGPM $\eta = 0.21(2)$ and $x_{1}^{\mathrm{typ}} = 0.14(2)$, meaning that both results are almost agree within statistical error bars. 

\subsection{Multifractility of the log-CFT}
Going beyond the typical scaling exponent, it is also interesting to probe the multifractal properties of the correlation function. At the critical point, the system is scale-invariant and all the moments of the correlation functions of the log-CFT~\cite{Aidan_PRL,LUDWIG_Multifractal,CHATELAIN_multifractal},
$C^{\vec{m}} = \mathrm{exp}\{t(\lambda_{1}^{\vec{m}}- \lambda_{0}^{\vec{m}})\}$ after averaging over trajectories (denoted by $\varepsilon$ ) decay as a power law in distance ($r$) in the long time limit,
\begin{equation}
\varepsilon [\{{C^{\vec{m}}}\}^n] \sim \frac{B_{n}}{r^{2x_{1}(n)}}.
\end{equation}
A cumulant expansion of the correlation function yields,
\begin{equation}
\ln[\varepsilon [\{{C^{\vec{m}}}\}^n] ] = n \varepsilon[\ln {C^{\vec{m}}}] + \frac{n^{2}}{2!} \varepsilon[\{ \ln  C^{\vec{m}} - \varepsilon[\ln {C^{\vec{m}}}]\}^2]+\dots
\label{eq:cumulants}
\end{equation}
This gives,
$x_{1}(n) = n x_{1}^{\rm{typ}} + \frac{n^{2}}{2!}x_{1}^{(2)} + \mathcal{O}(n^3)$ for sufficiently small $n$. We identify the presence of multi-fractality in the spectrum of correlation functions by the a non-zero estimate of $x_{1}^{(n)}$ where $ n \geq 2$. 

From the scaling of the second moment $k_2$ (taking $\alpha=1$)  defined as
\begin{equation}
\frac{k_2}{Lt} = \frac{\varepsilon\{ \ln  C^{\vec{m}} - \varepsilon[\ln {C^{\vec{m}}}]\}^2}{Lt} 
\end{equation}
and its slope vs $1/L^2$
we find a finite $x_1^{(2)}(\lambda=\Delta)=0.19(2)$ for the DGPM and $x_1^{(2)}(\Lambda=0.45)=0.14(2)$ for the SPMM. Thus our numerical results present evidence for multifractal scaling in MIPTs with weak measurements. These estimates are similar to that in the projective measurement case having $x_1^{(2)}=0.145(2)$ computed in {Ref.\cite{Aidan_PRL}}.

We further confirm the multifractal scaling by calculating the probability distribution function $P[Y(t)]$ of the correlation functions, where $Y(t) \equiv - \ln C^{\vec{m}}(t)$. In presence of multi-fractality, this is expected to follow the universal scaling form~\cite{LUDWIG_Multifractal,CHATELAIN_multifractal},
\begin{equation}
P[Y(t)] \sim \left(\frac{2\pi t}{L}\right)^{\frac{1}{2}}\exp\left[\frac{-2\pi t}{L} H\left(\frac{Y(t)}{2\pi t / L} \right) -b \right],
\label{eq:multifracscaling}
\end{equation}
collapsing onto a single curve $H(s)$ having a minimum at $s_o =x_{1}^{\mathrm{typ}}$, where we have choosen $b$ in Eq. \eqref{eq:multifracscaling} to bring $H(s_o)=0$.

 We  compute the distribution function $P[Y(t)]$ using data from systems sizes $L=8$ to $18$ and times $t=5L$ and $32L$ to find the universal scaling function $H(s)$ for the weak measurement models through the scaling form given in Eq. \ref{eq:multifracscaling}. $H(s)$ vs $s-s_o$ are shown in Fig.\ref{fig:LyapunovDGPM}(c) for the DGPM and in Fig.\ref{fig:Lyapunov_soft}(c) for the SPMM. Darker colors denote larger system sizes. We also plot the curve for projective measurement case by black dashed lines extracting the data points from Ref. \cite{Aidan_PRL} and shifting the origin of the $s$ axis to $s_o=x_1^{\mathrm{typ}}$. In both the models of weak measurements, the curves corresponding to weak measurements have similar shape compared to that of the projective measurement case, though not precisely overlapping. This may possibly arise from differences in higher $x_1^{(n)}$s in the multifractal spectra.

\begin{table}[h!]
\begin{tabular}{||c|c|c|c||} 
 \hline
  & $\frac{\lambda}{\Delta} = \infty$ & $\frac{\lambda}{\Delta} = 1$ &  $\Lambda = 0.45$ \\
 \hline\hline
 $p_{c}$ & 0.14(1) & 0.19(1) & 0.28(2) \\
\hline
 $\nu$ & 1.3(3) & 1.3(3) &  1.6(3)  \\
 \hline
 $z$ & 0.98(8) & 0.94(6) & 0.95(5) \\
 \hline
 $\eta$ & 0.23(2) & 0.21(2) & 0.19(3) \\
 \hline
 $c_{\mathrm{eff}}$ & 0.24(2) & 0.25(3) &  0.26(2) \\
  \hline
 $x_{1}^{\mathrm{typ}}$ & 0.12(2) & 0.14(2) & 0.12(2) \\
 \hline
 $x_{1}^{(2)}$ & 0.14(2) & 0.19(2) & 0.14(2) \\
 \hline\hline
\end{tabular}
\caption{\label{tab:table1} Critical data of universal parameters for the various models: central effective charge $c_{\mathrm{eff}}$, scaling dimension of the order paramter $x_{1}^{\mathrm{typ}}$, the multifractal critical exponent $x_{1}^{(2)}$, and the dynamical exponent $z$. These exponents were extracted at the critical probability $p_{c}$. At $\frac{\lambda}{\Delta} = 1$, the quantities involving the free energy ($c_{\mathrm{eff}}$, $x_{1}^{\mathrm{typ}},x_{1}^{(2)}$) were obtained from the DGPM, whereas the quantities involving the ancilla entanglement entropy ($\nu$,$z$,$\eta$) were obtained from the CGPM. Data is shown for the projective limit $\frac{\lambda}{\Delta} = \infty$, which is equivalent to $\Lambda = 1$.}
\end{table}

\section{Conclusion}\label{conclusion}
In this work, we have studied the effect of weak measurements on the critical properties of MIPTs. The summary of our findings are provided in Table~\ref{tab:table1}.
Based on scaling collapse we estimate $p_c$, $\nu$, $\eta$, and $z$, which reveals a Lorentz invariant transition with exponents 
 that are consistent with their values in the limit of strong projective measurements. As prior work found it challenging to determine the universality class from these estimates alone~\cite{Zabalo-2020} we turned to a more accurate transfer matrix approach. 
This is also based on the Lorentz invariance of the transition and the log-CFT nature of the field theory, allowing us to utilize universal finite size scaling corrections  to estimate  the effective central charge $c_{\mathrm{eff}}$  and the scaling dimension of the order parameter $x_1^{\mathrm{typ}}=\eta/2$, which both agree quite well across each model considered. We therefore take the agreement in $c_{\mathrm{eff}}$ and $x_1^{\mathrm{typ}}$ across the different measurement protocols as  the strongest evidence that each of  these models belong to the same universality class, namely the Haar random MIPT~\cite{Aidan_PRL}. 
We find clear evidence of multifractal scaling, which is indicitave of strong quantum fluctuations at the MIPT, in both measurement models. As these moments are higher statistical quantities they also carry much more numerical uncertainty. Therefore the larger discrepancies across models that we have observed in the multifractal properties, namely $x_1^{(2)}$ and the shape of $H(s)$, are not sufficient for us to make any firm conclusions regarding their higher moments differing. However, we leave this possibility open for future work, which will require orders of magnitude more statistical samples than we have considered here. $x_1^{(2)}$ for $\lambda/\Delta = 1$ appears to have a non-overlapping error bar when compared to other models. Similarly,  $x_1^{typ} = \eta/2$ for $\lambda/\Delta = 1$ does not agree within the error bars, although in this case the discrepancy is relatively minor.
An interesting and open question we leave for later is a construction of the log-CFT for the continuous Gaussian pointer model.
To summarize, taking all of numerical results together lead us to conclude that the nature of the universality class of the Haar MIPT and its underlying log-CFT description is unaffected by weak or strong measurement protocols.

\acknowledgements{
We thank Piers Coleman, Haining Pan, and Romain Vasseur for insightful discussions.
This work was partially supported by the Office of Naval Research grant No.~N00014-23-1-2357 (K.A., J.H.P.), the Army Research Office Grant No.~W911NF-23-1-0144 (K.A., J.H.P.), a Sloan Research Fellowship (J.H.P.), and the Abrahams Postdoctoral Fellowship at the Center for Materials Theory Rutgers (A.C.). We acknowledge the Office of Advanced Research Computing (OARC) at Rutgers, The State University of New Jersey for providing access to the Amarel cluster and associated research computing resources that have contributed to the results reported here as well as the services provided by the OSG Consortium, which is supported by the National Science Foundation awards No. 2030508 and No. 1836650.

\appendix
\begin{figure*}[t!]
\includegraphics[scale = 0.5]{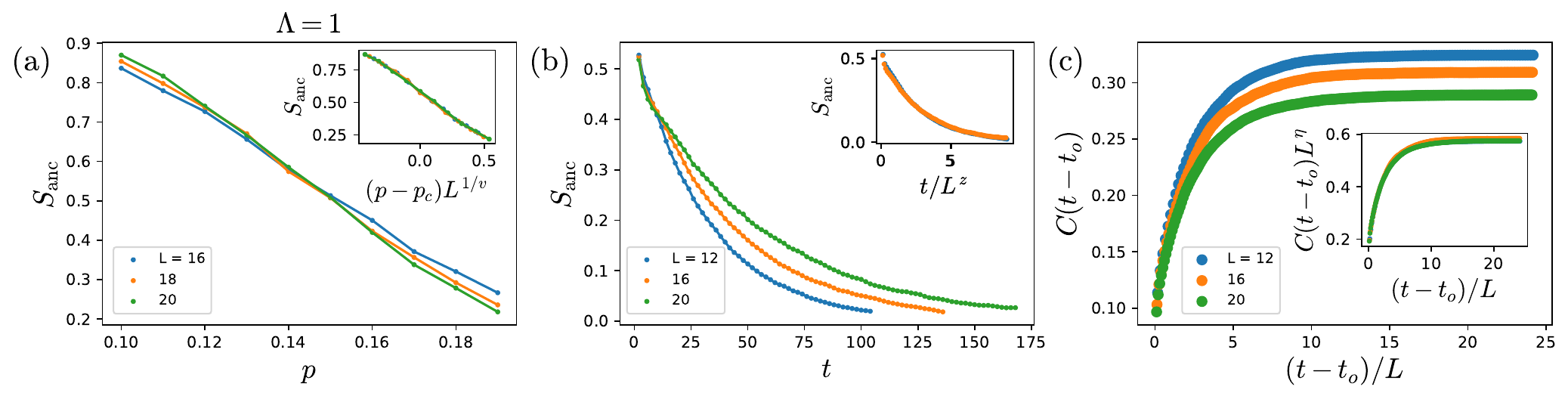}
\caption{{\bf Critical Properties of Strong Projective Measurements with HDU gates:} (a) Data is taken for 2,000 trajectories per system size. We find a crossing near $p=p_{c} = 0.14(1)$ 
consistent with previous work~\cite{Aidan_PRL}, and a scaling collapse of the data in the inset yielding $p_{c} = 0.14(1)$ and $\nu = 1.3(3)$. (b) Data is taken for 10,000 trajectories per system size. We find $z = 0.98(8)$ from a scaling collapse of the data, as shown in the inset. (c) Mutual Information between two ancilla qubits entangled at $r=1$ and $r'=\frac{L}{2}$ in the steady state of the circuit evolving with projective measurements. We use three different system sizes $L=12, 16, 20$ for the scaling collapse following Eq.~\eqref{eq:corrlcollapse} shown in the inset. We find the bulk exponent $\eta = 0.23(2)$. We average over $10^4$ trajectories for each system size.}
\label{fig:HDU}
\end{figure*}
\section{Critical properties of the Haar-dual-unitary Strong Projective Hybrid Quantum Circuit}
\label{sec:HDU}
In this Appendix we compute the remaining (easily accessible) critical exponents of the strongly projective HDU MIPT. In Ref.~\cite{Aidan_PRL} $p_c$ and $x_1^{\mathrm{typ}}$ were computed. Here we compute $\nu$ and $\eta$ from the ancilla probes in Sec.~\ref{sec:phasediagram} in order to further verify that the HDU gates do not affect the universal nature of the MIPT. As shown in Fig.~\ref{fig:HDU}(a) and (c) we find that, for the von Neumman entropies, $\eta^{\rm{HDU}} = 0.23(2)$, and $\nu^{\rm{HDU}} = 1.3(3)$ for HDU gates with strong measurement, which are close to the values $\eta^{\rm{Haar}}=0.19(1)$ and $\nu^{\rm{Haar}} = 1.2(2)$ for Haar random gates with strong measurement as computed in Ref.\onlinecite{Zabalo-2020}. Fig.~\ref{fig:HDU}(b) also shows signatures of Lorentz invariance in HDU circuit.
In conclusion, we find good agreement between HDU and Haar random gates as expected.

\section{Comparison of Discrete and Continuous Reyni Entropies}
\label{sec:AppendixB}
In this section, we compare the Reyni entanglement entropies in the discrete and continuous Gaussian weak measurement models. We compute both the ancilla and bipartite entanglement entropies following the protocols explained in Secs.~\ref{orderparameter} and \ref{probhalfcut}. We find that for $L=12$ and all measurement probabilities centered around the critical point $p_{c} \sim 0.19$, the Reyni entropies in both models are virtually indistinguishable. This allows us to use either measurement model to compute the Reyni entropies. In Fig. \ref{ctsdiscomp}, we compute the bipartite entanglement entropies in (a) and the ancilla entanglement entropies in (b) for both the CGPM and DGPM.
The difference in entanglement entropies is orders of magnitude smaller than the mean value in both instances: $\frac{\Delta{S_{\mathrm{L/2}}}}{S_{\mathrm{L/2}}} \sim 10^{-2}$ and $\frac{\Delta{S_{\mathrm{anc}}}}{S_{\mathrm{anc}}} \sim 10^{-3}$. The difference is also of the same order of magnitude as the statistical error bars in both models, which is smaller than the size of the points.

\section {Krauss Operators for Weak Measurement}
\label{sec:AppendixC}
For DGPM, we showed that $p(x_{o};\lambda/\Delta,\epsilon)$ could be defined in terms of the integrals $p_{\pm}(x_{o},\lambda/\Delta,\epsilon)$, as expressed in Eq.~\eqref{eq:diskrauss}. In this appendix we describe how we numerically evaluate the Born probabilities $p_{\pm}(x_o;\lambda/\Delta,\epsilon)$ defined in Eq.~\eqref{eq:diskrauss}. To numerically compute these integrals we consider the limit $\epsilon \rightarrow 0$. In this limit,  the integral over position can be efficiently calculated by numerical evaluation of the error function:
\begin{equation}
\mathrm{erf}(x)=\frac{2}{\sqrt{\pi}}\int_{0}^{x}e^{-t^2}dt.
\end{equation}
In terms of the error function the probabilities become,
\begin{equation}
\begin{split}
    p_{\pm}(x_{o};\lambda/\Delta,\epsilon)= & \frac{\sqrt{\Delta}}{2}\Big[\mathrm{erf}\left(\frac{\mp 2\lambda -2x_{o}+\epsilon}{2 \Delta}\right) 
    \\ & - \mathrm{erf}\left(\frac{\mp 2\lambda -2x_{o}-\epsilon}{2 \Delta}\right)\Big]
\end{split}
\label{DGPM_analytical}
\end{equation}
Therefore, the Krauss operator which updates the state, as represented in Eq.\eqref{eq:updatediscrete}, becomes
\begin{equation}
\begin{split}
    \hat{M}_{D}(x_{o}) = & \frac{1}{2 \sqrt{\Delta}}\Big\{\Big[\mathrm{erf}\left(\frac{-2x_{o}-2\lambda + \epsilon}{2\Delta}\right)
    \\ & - \mathrm{erf}\left(\frac{-2x_{o}-2\lambda - \epsilon}{2\Delta}\right)\Big] \Pi_{+}^{(j)}
    \\ & + \Big[\mathrm{erf}\left(\frac{-2x_{o}+2\lambda + \epsilon}{2\Delta}\right)
    \\ & - \mathrm{erf}\left(\frac{-2x_{o}+2\lambda - \epsilon}{2\Delta}\right)\Big] \Pi_{-}^{(j)} \Big\}.
\end{split}
\end{equation}
\begin{figure}[t!]
\includegraphics[scale = 0.35]{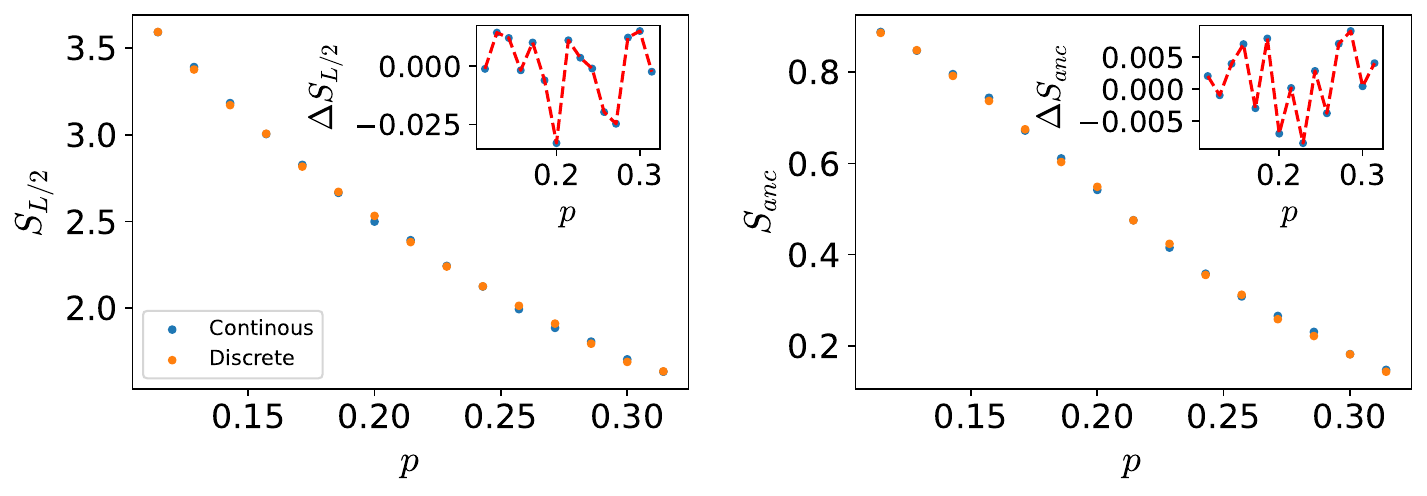}
\caption{{\bf Comparison of Reyni Entropies}. Data is taken for $L = 12$ and 2,000 trajectories per data point, at measurement strength $\frac{\lambda}{\Delta} = 1$, $\epsilon = 0.1$ and $\Delta = 10,000$ for the discrete model, and $\frac{\lambda}{\Delta} = 1$ for the continous measurement model. (a) Comparison of the bipartite entanglement entropies in the continuous and discrete Gaussian measurement models. The inset is the difference in the computed entangled entropies $\Delta S_{L/2} = S_{\mathrm{L/2}}^{(\mathrm{Discrete})}-S_{\mathrm{L/2}}^{(\mathrm{Continous})}$(b) Comparison of the ancilla entanglement entropies in the continuous and discrete Gaussian measurement models. The inset presents $\Delta S_{\mathrm{anc}} = S_{\mathrm{anc}}^{(\mathrm{Discrete})}-S_{\mathrm{anc}}^{(\mathrm{Continous})}$.
}
\label{ctsdiscomp}
\end{figure}
\section{Changing  $\epsilon$ does not affect $c_{\mathrm{eff}}$}\label{app:ceff}
In this appendix we show that for DGPM, the effective central charge we calculated is independent of the choice of the binning width $\epsilon$. 
In the limit $\epsilon \ll \Delta$ (there are no constraints on $\lambda$, except noting that for a fixed strength $\lambda/\Delta$, $\lambda$ depends on $\Delta$) leading us to show in this appendix that
the change in the free energy as a function of the width of the measurement region is given by
\begin{equation}
    \Delta f = f(\epsilon') - f(\epsilon) =  -p_{c}\mathrm{log}\left(\frac{\epsilon}{\epsilon'}\right),
    \label{eqn:deltaf}
\end{equation}
where $\epsilon$ and $\epsilon'$ are two different bin widths.
Consequently, altering $\epsilon'$ in this limit leaves $c_{\mathrm{eff}}$ invariant, as an $L$-independent shift of $f$ only changes $f(L=\infty)$. 

We now provide a derivation of Eq.~\eqref{eqn:deltaf}.In the limit $\epsilon \ll \Delta$, the integrals in Eq.\eqref{eq:diskrauss} simplify to become
\begin{equation}
    p_{\pm}(x_{o};\lambda/\Delta,\epsilon)= G_{\Delta}^{2}(x'\mp \lambda  )\epsilon.
\end{equation}
Therefore the probability in Eq. \eqref{eq:pxoDiscrete-main} becomes 
\begin{eqnarray}
    p(x_{o};\lambda/\Delta,\epsilon)&=&[\langle \psi|\Pi_{+}^{(j)}| \psi \rangle G_{\Delta}^{2}(x_{o}- \lambda) 
    \nonumber
    \\
    &+& \langle \psi|\Pi_{-}^{(j)}| \psi \rangle G_{\Delta}^{2}(x_{o} + \lambda)]\epsilon.
    \label{eq:pxoDiscrete}
\end{eqnarray} 
We use $p(x_{o}^{(i)}; \lambda/\Delta, \epsilon)$ to denote the probability of the $i^{th}$ measurement outcome in the DGPM,
and $p(x_{o}^{(i)};\lambda/\Delta)$ to denote the corresponding probability density in the CGPM.
Using $N_{\mathrm{meas}}$ to indicate the total number of measurements in the circuit, we can now express the probability of the measurement record in the trajectory $p_{\vec{m}}$ as: 
\begin{equation}
\begin{aligned}
p_{\vec{m}_{\epsilon}} &= \prod_{i=1}^{N_{\text{meas}}} p(x_{o}^{(i)}; \lambda/\Delta, \epsilon) \\
&= \epsilon^{N_{\text{meas}}} \prod_{i=1}^{N_{\text{meas}}} p(x_{o}^{(i)};\lambda/\Delta).
\end{aligned}
\label{eq:poDiscreteap}
\end{equation}
We now use $\vec{m}_{\epsilon}$ to denote each possible measurement trajectory at the corresponding bin width $\epsilon$. We express 
\begin{equation}
\vec{m}_{\epsilon}=(x_{o}^1(\epsilon),x_{o}^2(\epsilon),x_{o}^3(\epsilon),\dots)  
\end{equation}
where $x_{o}^i(\epsilon$) refers to the measured pointer position at the $i^{th}$ measurement event, which can depend on $\epsilon$.
For each measurement we use $N_{\mathrm{out}}(\epsilon)$ to denote the number of measurement outcomes for $x_{o}^{i}(\epsilon)$ in a finite spatial region of width $W$. We see that 
the number of outcomes $N_{\mathrm{out}}$ depends on the bin width $\epsilon$ as 
\begin{equation}
N_{\mathrm{out}}(\epsilon) = \frac{W}{\epsilon}.
\end{equation}
For example, halving the bin width doubles the number of measurement outcomes over any region.
Therefore the number of measurement trajectories $\vec{m}_{\epsilon}$ will scale as
\begin{equation}
N_{\mathrm{traj}}(\epsilon) = \left(\frac{W}{\epsilon}\right)^{N_{\mathrm{meas}}}.
\end{equation}
This leads to the observation that in the limit $\epsilon \ll \Delta$, the sum over measurement trajectories depends on the width of the measurement region as: 
\begin{equation}
\sum_{\vec{m}_{\epsilon}}= \left( \frac{\epsilon'}{\epsilon} \right)^{N_{\text{meas}}} \sum_{\vec{m}_{\epsilon'}}.
\label{eq:region}
\end{equation} 
Now, recall that we can write the free energy as
\begin{equation}
F_{\epsilon} = -\sum_{\vec{m}_{\epsilon}}p_{\vec{m}_{\epsilon}}\mathrm{ln}p_{\vec{m}_{\epsilon}}.
\end{equation}
$F_{\epsilon}$ represents the free energy using a bin width $\epsilon$. Using Eq. \eqref{eq:poDiscreteap} to relate $p_{\vec{m}_{\epsilon}}$ and $p_{\vec{m}_{\epsilon'}}$, we have that
\begin{equation}
\begin{aligned}
F_{\epsilon} & = -\sum_{\vec{m}_{\epsilon}} p_{\vec{m}_{\epsilon}}\mathrm{ln}\left[p_{\vec{m}_{\epsilon'}}\left(\frac{\epsilon}{\epsilon'}\right)^{N_{\text{meas}}}\right] \\ 
& = -\sum_{\vec{m}_{\epsilon'}} p_{\vec{m}_{\epsilon'}}\mathrm{ln}p_{\vec{m}_{\epsilon'}} -\sum_{\vec{m}_{\epsilon}} p_{\vec{m}_{\epsilon}}N_{\mathrm{meas}}\mathrm{ln}\left(\frac{\epsilon}{\epsilon'}\right) \\
& = F_{\epsilon'} -N_{\text{meas}}\ln\left(\frac{\epsilon}{\epsilon'}\right) .
\end{aligned}
\label{eq:poDiscretea2}
\end{equation}
We now take $\varepsilon[...]$ to denote an average over circuits, so that at $p=p_{c}$ we expect that $\varepsilon[N_{meas}] \approx tLp_{c}$. Thus,
\begin{eqnarray}
\Delta F & =  \varepsilon[F_{\epsilon}] - \varepsilon[F_{\epsilon'}]
         \\ 
         & = -Ltp_{c}\mathrm{ln}\left(\frac{\epsilon}{\epsilon'}\right)
\end{eqnarray}
and 
\begin{equation}
\Delta f = \frac{\Delta F}{Lt} = -p_{c}\mathrm{ln}\left(\frac{\epsilon}{\epsilon'}\right).
\end{equation}
$\Delta f$ represents the change in $f(L=\infty)$ as $\epsilon$ is varied, which is an $L$-independent shift to the free energies at all system sizes at a fixed measurement strength.
As shown in Fig. \ref{fig:Feps}, we perform a calculation of the free energy for various values of $\epsilon$, at the measurement strength $\frac{\lambda}{\Delta} = 1$. We verify that our analytical calculation for the dependence of $f$ on $\epsilon$ matches the numerics. 
\begin{figure}[ht]
\includegraphics[scale = 0.6]{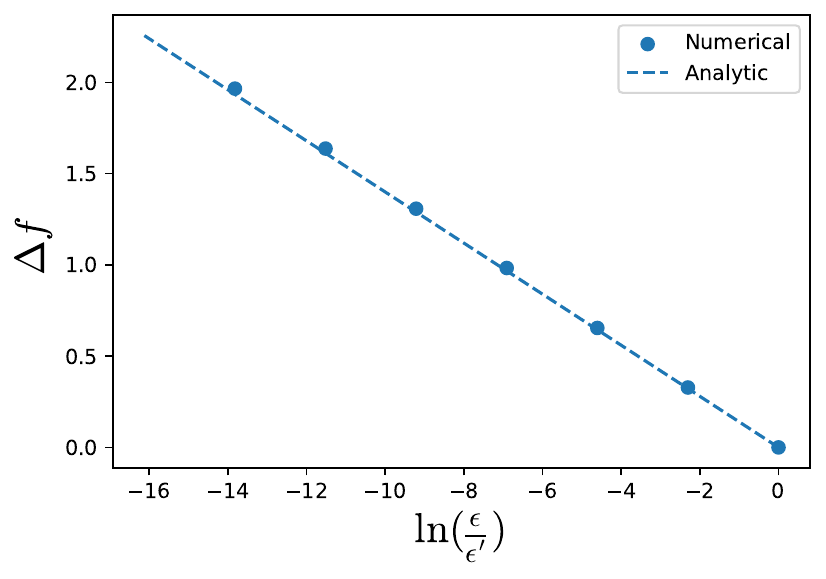}
\caption{{\bf Free Energy vs. $\epsilon$}. (a) Dependence of $f$ on $\epsilon$ in the Gaussian weak measurement model at $p \sim p_{c} = 0.19$ and $\frac{\lambda}{\Delta} = 1$ with $\Delta = 1,000,000$ and $\epsilon = 0.1$ and $\epsilon' = 1$.}
\label{fig:Feps}
\end{figure}
\newpage
\bibliography{References.bib}

\end{document}